\documentclass[3p]{elsarticle}
\usepackage{graphicx}
\usepackage{amsmath}
\usepackage{amssymb}

\usepackage{color}

\newcommand{\sph}[1]{\left\langle #1 \right\rangle}
\newcommand{\sphint}[1]{\widehat{#1}}
\newcommand{\q}{q}
\newcommand{\Reynolds}{\textit{Re}}
\newcommand{\mean}[1]{\overline{#1}}

\title{Analysis of the incompressibility constraint in the Smoothed Particle Hydrodynamics method }

\author[imp]{K.~Szewc\corref{cor1}}
\ead{kszewc@imp.gda.pl}

\author[imp]{J.~Pozorski}
\ead{jp@imp.gda.pl}

\author[edf]{J.-P.~Minier}

\cortext[cor1]{Corresponding author: kszewc@imp.gda.pl, tel. +48 58 6995238,
fax. +48 58 3416144}

\address[imp]{Institute of Fluid-Flow Machinery, Polish Academy of Sciences, 
ul. Fiszera 14, 80-952 Gda\'nsk, Poland}

\address[edf]{Electricit\'e de France, Direction de la Recherche et du D\'eveloppement,
D\'epartement M\'ecanique de Fluides, Energies et Environnement, 6 quai Watier, F-78401 Chatou, France}

\begin{document}

\begin{abstract}
Smoothed particle hydrodynamics is a particle-based, fully Lagrangian, method for fluid-flow simulations.
In this work, fundamental concepts of the method are first briefly recalled. 
Then, we present a thorough comparison of three different incompressibility treatments in SPH:
the weakly compressible approach, where a suitably-chosen equation of state is used;
and two truly incompressible methods, where the velocity field projection onto a divergence-free space is performed.
A noteworthy aspect of the study is that, in each incompressibility treatment, 
the same boundary conditions are used (and further developed)
which allows a direct comparison to be made. Problems associated with implementation are also discussed and an optimal
choice of the computational parameters has been proposed and verified. Numerical results show that the present
state-of-the-art truly incompressible method (based on a velocity correction) suffer from density accumulation errors.
To address this issue, an algorithm, based on a correction for both particle velocities and positions, is presented.
The usefulness of this density correction is examined and demonstrated in the last part of the paper.
\end{abstract}


\maketitle

\section{Introduction}
\label{sec:introduction}
Smoothed Particle Hydrodynamics (SPH) is a fully Lagrangian, particle-based
approach for fluid-flow computations. This method was independently proposed by
Gingold \& Monaghan \cite{Gingold & Monaghan 1977} and Lucy \cite{Lucy 1977} to
simulate some astrophysical phenomena at the hydrodynamic level (compressible
flow). Nowadays, SPH is more and more often used
for flows with interfaces and in geophysical applications. Its
main advantage over Eulerian techniques is that there is no requirement of
the grid. Therefore, the SPH method is natural to use for complex
geometries or multi-phase flows. 

An important issue in SPH is the proper
implementation of the incompressibility constraints. In the present work, three
different implementations are critically discussed and compared for a selection
of validation examples. The first one is the Weakly Compressible SPH
(WCSPH), which is the most common technique. It involves the standard
set of
governing equations closed by a suitably-chosen, artificial equation of state,
cf. Sect.~\ref{sec:weakly compressible sph}. The second and third
implementations, called truly incompressible SPH (ISPH), are based on the
Projection Method, introduced in SPH by Cummins \& Rudman \cite{Cummins & Rudman 1999}.
Generally in ISPH, a Poisson equation is solved to obtain the divergence-free
velocity field, cf. Sect.~\ref{sec:truly incompressible sph}. 
In the second approach, this Poisson equation is solved on an auxiliary regular mesh.
Here, this method is called the Grid-Projected Poisson Solver (GPPS). 
Since this approach consists in combining the Eulerian and Lagrangian techniques, it is not an optimal choice for free-surface flows; yet, since SPH has a huge potential in the subject of multi-phase flows, its usefulness seems to be worth of discussion.
The third approach considered is
the ISPH scheme with the Particle Poisson Solver (PPS), cf. Sect.~\ref{sec:particle
poisson solver}. In this concept the Poisson equation is rewritten in SPH
formulation and solved on the grid of Lagrangian points (particles). The main
advantage of this idea is that the additional domain discretization and related
discretization errors are avoided. The paper is also related to the work of Lee et al.~\cite{Lee et al. 2008}, 
where a similar comparison between WCSPH and ISPH -
PPS approaches has been done. However, in the present work, a comprehensive
analysis of boundary conditions is performed.
Moreover, since we use the same (ghost-particle) boundary conditions for both
WCSPH and ISPH approaches, our comparison offers a new look at the usefulness of
the incompressibility variants. 
For the sake of validation for both single- and two-phase flows, the three
different implementations of the incompressibility constraints are compared for
the lid-driven cavity flow and the Rayleigh-Taylor
instability. For the former case, the impact of various computational
parameters on the results have been analyzed and optimal setting has been found;
it is consequently used also for the latter case.

A very good comparison of ISPH solvers was presented by Xu et al.~\cite{Xu et al. 2009}.
However, the authors do not discuss the problem of accumulating density errors
in the ISPH approach.  
Since, like other authors \cite{Cummins & Rudman 1999} \cite{Hu & Adams 2007},
we have been faced with such a problem, we decided to examine the usefulness of the correction
algorithm proposed by Pozorski and Wawre\'nczuk \cite{Pozorski & Wawrenczuk 2002}, 
cf. Sect.~\ref{sec:pozorski and wawrenczuk constant-density approach}.
This formulation, used also in the PDF computations of turbulent flows
\cite{Minier & Pozorski 1999}, consists in retrieving the constant fluid
density field by applying the correction to the particles' position. In
practice, that procedure involves the computations of the second Poisson
solver. The effectiveness of such an approach and the improvement of results
are demonstrated.

Another common field of research in SPH is the proper implementation of the
boundaries. In the present work, among many techniques, the ghost-particle
boundary approach is applied. This implementation involves the use of fictitious
external particles that are reflections of the fluid particles in the
computational domain. The main advantages of this method are simplicity and
conformity with different phases of the fluid \cite{Valizadeh et al. 2008}. 
Computing the lid-driven cavity test problem with the WCSPH approach, we have
found that the standard implementation of the no-slip boundary condition by the
ghost-particle approach \cite{Cummins & Rudman 1999} may cause the stability
problems. Our proposal to overcome these difficulties is based on the
combination of the free-slip and no-slip boundary conditions.
The detailed discussion of this problem is presented in Sect.~\ref{sec:boundary
conditions}.

\section{Formulation of the SPH method}
\label{sec:formulation of the SPH method}
The main idea behind SPH is the introduction of the kernel interpolants
for the field quantities so that the fluid dynamics is represented by particle
evolution equations. The SPH method is based on three approximations.

The first is interpolation of the field quantities at a point. To construct it,
we utilize an integral interpolant $\sphint A(\mathbf r)$ of any field
$A(\mathbf r)$ (for simplicity we consider here a scalar field)
\begin{equation} \label{integral interpolant}
	\sphint A(\mathbf r) = \int_{\Omega} A(\mathbf r') W(\mathbf r - \mathbf r', h) d\mathbf r',
\end{equation}
where the integration is over all the domain $\Omega$ and $W(\mathbf r, h)$
is a weighting function (the kernel) with the parameter $h$ called the smoothing
length (a linear dimension of smoothing). Generally, the kernel should posses a
symmetrical form
\begin{equation} \label{kernel symmetry}
	W(\mathbf r, h) = W(-\mathbf r, h),
\end{equation}
satisfy the limit condition
\begin{equation} \label{kernel prop 1}
	\lim_{h \rightarrow 0} W(\mathbf r, h) = \delta(\mathbf r),
\end{equation}
where $\delta({\mathbf r})$ is the Dirac delta distribution, and should be
normalized so that
\begin{equation} \label{kernel prop 2}
	\int_{\Omega} W(\mathbf r, h) d\mathbf r = 1.
\end{equation}
Additionally, the kernel should be at least as many times differentiable as the field $A$.
Taking into consideration the computational effort and the proper implementation of the boundary conditions (Sect.~\ref{sec:boundary conditions}), it is worth to use compact support kernels. Since there are a lot of possibilities to choose $W(\mathbf r, h)$, we decided to compare solutions obtained employing three different 2D kernels (Sect.~\ref{sec:lid-driven cavity}): 
the cubic B-spline form
\begin{equation} \label{cubic spline kernel}
 W(\mathbf r, h) = \frac{10}{7\pi h^2} \begin{cases}
       1 - \frac{3}{2} \q^2 + \frac{3}{4} \q^3, & \text{for } 0 \leq \q \leq 1, \\
       \frac{1}{4} \left(2-\q\right)^3, & \text{for } 1 \leq \q \leq 2, \\
       0, & \text{otherwise},
        \end{cases}
\end{equation}
the quintic form proposed by Wendland \cite{Wendland 1995}
\begin{equation} \label{quintic Wendland}
W(\mathbf r, h) = \frac{7}{4\pi h^2}  \begin{cases}
    \left( 1-\frac{q}{2} \right)^4
\left(2q+1\right), & \text{for } |\mathbf r| \leq 2h, \\
    0, & \text{otherwise},
  \end{cases}
\end{equation}
and the quintic B-spline form introduced by Morris \cite{Morris et al. 1997}
\begin{equation} \label{quintic Morris}
W(\mathbf r, h) = \frac{7}{478\pi h^2} \begin{cases}
	\left(3-\q \right)^5 - 6\left(2-\q \right)^5 + 15\left(1-\q \right)^5, & \text{for } 0 \leq \q \leq 1, \\
	\left(3-\q \right)^5 - 6\left(2-\q \right)^5, & \text{for } 1 \leq \q \leq 2, \\
	\left(3-\q \right)^5, & \text{for } 2 \leq \q \leq 3, \\
	0, & \text{otherwise},
\end{cases}
\end{equation}
where $q = |\mathbf r|/h$.

The second approximation of the SPH technique is the discretization of space. It
is done through dividing the domain into a fine-grained representation
(particles). Each particle carries the properties of the field. The integral
interpolant $\sphint{(\cdot)}$, Eq.~(\ref{integral interpolant}), becomes then
the summation interpolant $\sph{\cdot}$
\begin{equation} \label{summation interpolant}
\sph{A}(\mathbf r) = \sum_{b} A(\mathbf r_b) W(\mathbf r - \mathbf r_b, h)
\Omega_b,
\end{equation}
where $\mathbf r_b$ and $\Omega_b$ denote the position and volume of the
particle $b$. The SPH task involves the foregoing computations of the
interpolant at each particle, so that Eq.~(\ref{summation interpolant}) may be
rewritten into the common form
\begin{equation} \label{SPH interpolant}
\sph A_a = \sum_b A_b W_{ab}(h) \Omega_b,
\end{equation}
where $\sph{A}_a = \sph{A}(\mathbf r_a)$, $A_a = A(\mathbf r_a)$ and $W_{ab}(h) = W_{ba}(h) = W(\mathbf r_b - \mathbf r_a, h)$.

An additional advantage of SPH reveals with the differentiation of fields.
In accordance with (\ref{integral interpolant}), the gradient of $A(\mathbf r)$
assumes the form
\begin{equation} \label{nabla integral interpolant}
\sphint{\nabla A}(\mathbf r) = \int_{\Omega} \nabla A(\mathbf r') W (\mathbf r - \mathbf r', h) d\mathbf r'.
\end{equation}
Taking advantage of the integration by parts rule and utilizing the kernel symmetry, we can transform the foregoing equation into
\begin{equation} \label{nabla integral interpolant after integration by parts}
\sphint{\nabla A}(\mathbf r) = \int_{\partial \Omega} A(\mathbf r')W(\mathbf r - \mathbf r') \mathbf n' dS + \int_{\Omega} A(\mathbf r') \nabla' W (\mathbf r - \mathbf r', h) d\mathbf r',
\end{equation}
where $\mathbf n'=\mathbf n({\mathbf r'})$ is the normal vector to surface $\partial \Omega$.
Generally, the first term does not necessarily vanish for finite domain sizes, cf.~\cite{Feldman & Bonet 2007}.
However, it is a common practice to neglect this term and deal with the boundaries
explicitly. The SPH form (discretization) of (\ref{nabla integral interpolant
after integration by parts}) brings the common rule
\begin{equation} \label{SPH nabla interpolant}
\sph{\nabla A}_a = \sum_b A_b \nabla_a W_{ab}(h) \Omega_b.
\end{equation}
Since the nabla operator acts only on the kernel, the gradient of the field is dependent only on the values of the fields at particles, not gradients.

The third SPH feature is the assumption that the field value $A_a$ at a point
and its SPH approximation $\sph{A}$ are in relation
\begin{equation}
\sph{A}_a \approx A_a.
\end{equation}
Since, in the case of the time dependent simulations, for each time step, the calculation of the field quantities is performed using the results obtained in the previous step, the above introduced assumption becomes natural.

\section{Governing equations}
\label{sec:governing equations}
The full set of governing equations for incompressible viscous flow is composed
of the Navier-Stokes (N-S) equation
\begin{equation} \label{NS}
\frac{d \mathbf u}{dt} = -\frac{1}{\varrho} \nabla p + \nu \nabla^2 \mathbf u +
\mathbf f,
\end{equation}
where $\varrho=const$ is the density, $\mathbf u$ the velocity, $t$ the time, $p$ the pressure, $\nu$ the kinematic viscosity and $\mathbf f$ an external force, and the continuity equation
\begin{equation} \label{continuity equation}
\frac{d\varrho}{dt} = -\varrho \nabla \cdot \mathbf u,
\end{equation}
that for $\varrho = const$ arises to the form
\begin{equation} \label{div u = 0}
\nabla \cdot \mathbf u = 0.
\end{equation}

The whole set of governing equations should be expressed in the SPH formalism.
Utilizing the relation (\ref{SPH nabla interpolant}), the divergence of velocity
becomes
\begin{equation} \label{SPH div u basic}
\sph{\nabla \cdot \mathbf u}_a = \sum_{b} \mathbf u_b \cdot \nabla_a
W_{ab}(h)\Omega_b.
\end{equation}
Therefore, the continuity equation (\ref{continuity equation}) takes the form
\begin{equation} \label{SPH continuity equation basic}
\frac{d\varrho_a}{dt} = -\varrho_a \sum_b \frac{m_b}{\varrho_b} \mathbf{u}_b \cdot \nabla_a W_{ab}(h),
\end{equation}
where $m_b/\varrho_b = \Omega_b$.
It is important to note that various ways exist to express the divergence, for
example, 
\begin{equation} \label{SPH continuity symmetrical}
\frac{d\varrho_a}{dt} = \sum_{b} m_b \mathbf{u}_{ab} \cdot \nabla_a W_{ab}(h),
\end{equation}
where $\mathbf{u}_{ab} = \mathbf u_a - \mathbf u_b$.
The advantage of the above form over (\ref{SPH continuity equation basic}) is the symmetry with swapping particles $a$ and $b$. Therefore, in practice, it is more accurate to use (\ref{SPH continuity symmetrical}) \cite{Morris 1996}. However, there exists an alternative formulation. The fluid density can be computed directly from the SPH formula (\ref{SPH interpolant})
\begin{equation} \label{SPH direct density computation}
\varrho_a = \sum_b \varrho_b W_{ab}(h) \Omega_{b} = \sum_b m_b W_{ab}(h).
\end{equation}
A practical disadvantage of this approach is that $\varrho$ must be evaluated by
summing over the particles before other quantities \cite{Morris et al. 1997}.
Therefore, the computational effort increases (Sect.~\ref{sec:lid-driven
cavity}). Another disadvantage is the problem with representing sharp
discontinuities near material interfaces. To avoid this difficulty, Hu \&
Adams \cite{Hu & Adams 2006} suggested to use the form
\begin{equation} \label{SPH direct density computation multiphase}
\varrho_a = m_a \sum_b W_{ab}(h).
\end{equation}

In the SPH technique,
the pressure N-S term is responsible for ensuring the incompressibility
constraint
(Sect.~\ref{sec:incompressibility treatment}). Utilizing (\ref{SPH nabla
interpolant}) it takes the form
\begin{equation} \label{SPH NS pressure term basic}
\sph{\frac{1}{\varrho} \nabla p}_a = \frac{1}{\varrho_a} \sum_b \frac{m_b}{\varrho_b} p_b \nabla_a W_{ab}(h).
\end{equation}
Similarly to the continuity equation (\ref{SPH continuity symmetrical}) there is a possibility to obtain more useful gradient approximations. In the present paper, we utilize the form proposed by Colagrossi and Landrini \cite{Colagrossi & Landrini 2003}
\begin{equation} \label{SPH NS pressure term symmetrical colagrossi}
\sph{\frac{1}{\varrho} \nabla p}_a = \sum_{b} m_b \frac{p_a + p_b}{\varrho_a \varrho_b} \nabla_a W_{ab}(h).
\end{equation}
 is expressed as a combination of the finite difference and SPH discretisations \cite{Cummins et al. 1997} \cite{Cleary & Monaghan 1999}.
For the present work, we utilize the form (with a small regularising parameter 
$\eta=0.01h$) \cite{Cleary & Monaghan 1999}
\begin{equation} \label{SPH NS viscous term}
\sph{\nabla (\nu \nabla \cdot \mathbf u)}_a = \sum_{b} m_b \left( 8 \frac{\nu_a
+ \nu_b}{\varrho_a + \varrho_b} \frac{\mathbf u_{ab} \cdot \mathbf
r_{ab}}{r_{ab}^2 + \eta^2} \right) \nabla_a W_{ab}(h).
\end{equation}
Since SPH is fully Lagrangian approach, the particle advection equation
completes the system
\begin{equation} \label{SPH equation of motion}
\frac{d\mathbf r_a}{dt} = \mathbf u_a.
\end{equation}

\section{Incompressibility treatment}
\label{sec:incompressibility treatment}
To compute incompressible flows in the Eulerian CFD, two approaches are commonly
used: the artificial compressibility method (where a specific equation of state
is applied) and the pressure-correction technique (where the velocity field is
projected onto the divergence-free space). Generally, the analogs of these
techniques are also used in the SPH approach. They are presented in the
following.

\subsection{Weakly Compressible SPH}
\label{sec:weakly compressible sph}
The most common technique is the weakly compressible SPH (WCSPH). It involves
the set of governing equations closed by a suitably chosen, artificial equation
of state $p=p(\varrho)$. Since the fluid pressure is an explicit function of
$\varrho$, the density gradient exerts an influence on the particle motion. The
commonly used equation of state has the form \cite{Batchelor 1967}
\begin{equation} \label{equation of state}
p=\frac{c^2\varrho_0}{\gamma} \left[ \left( \frac{\varrho}{\varrho_0} \right)^\gamma - 1 \right],
\end{equation}
where the reference density $\varrho_0$, the numerical sound speed $c$ and
parameter $\gamma$ are suitably chosen to reduce the density fluctuations down
to demanded level. In the present work, to assure density variations lower than
1\%, we set $\gamma=7$ and $c$ at the level at least $10$ times higher than the
maximal fluid velocity~\cite{Monaghan 1994}. However, since the sound speed is high, the time step
should be, correspondingly, very small (cf. Sect.~\ref{sec:time step criteria}).
Therefore, the computational efficiency is the main weakness of the WCSPH method.

\subsection{Truly Incompressible SPH with the grid-projected Poisson solver}
\label{sec:truly incompressible sph}
The newest promising technique is the truly incompressible SPH (ISPH). It is
based on the Projection Method which is a common approach for numerically
solving time-dependent incompressible fluid-flow problems. In this technique the
pressure needed to ensure incompressibility is found by projecting the
calculated velocity field onto the divergence-free space \cite{Cummins & Rudman 1999}. 
This is possible due to the Helmholtz decomposition which states: every
vector field $\mathbf A$ that is twice continuously differentiable and vanishes
faster than $1/r$ at infinity can be decomposed into the gradient and the curl
parts as follows \cite{Gryffits 1999}
\begin{equation} \label{Helmholtz decomposition}
\mathbf A = \nabla \phi + \nabla \times \mathbf B = \mathbf A_{\text{curl free}} + \mathbf A_{\text{div free}},
\end{equation}
where $\phi$ and $\mathbf B$ are suitably chosen and
\begin{equation} \label{Helmholtz decomposition projection}
\begin{split}
\nabla \times \mathbf A_{\text{curl free}} &= \nabla \times (\nabla \phi) = 0,
\\
\nabla \cdot \mathbf A_{\text{div free}} &= \nabla \cdot (\nabla \times \mathbf
B) = 0.
\end{split}
\end{equation}
In the ISPH approach, the decomposition procedure begins with splitting
the integration of the N-S equation (\ref{NS}) on the time interval $\delta t=t^{n+1}-t^n$
into two parts. The first, so-called predictor step, gives the fractional
velocity $\mathbf u^*$
\begin{equation} \label{ISPH predictor}
\frac{\mathbf u^* - \mathbf u^n}{\delta t} = \nu \nabla^2 \mathbf u^n + \mathbf
f^n.
\end{equation}
The right-hand side of the above equation contains all the N-S terms except the
one connected with the pressure.
The second part of the procedure is the correction step
\begin{equation} \label{ISPH corrector}
\frac{\mathbf u^{n+1} - \mathbf u^*}{\delta t} = -\frac{1}{\varrho} \nabla p^{n+1}.
\end{equation}
It imposes the correction of $\mathbf u^*$ to ensure compliance with the
divergence-free constraint. To obtain an appropriate pressure
$p^{n+1}$ we write the divergence of Eq.~(\ref{ISPH corrector})
\begin{equation} \label{div of ISPH corrector}
\nabla \cdot \left(\frac{\mathbf u^{n+1} - \mathbf u^*}{\delta t} \right) =\nabla \cdot \left( -\frac{1}{\varrho} \nabla p^{n+1} \right).
\end{equation}
Since we expect a divergence-free velocity field at the end of the time step, we
require that $\nabla \cdot
\mathbf u^{n+1}=0$. Therefore, the formula (\ref{div of ISPH corrector}) arises
into the Poisson equation
\begin{equation} \label{Poisson}
\nabla \cdot \left( \frac{1}{\varrho} \nabla p^{n+1} \right) = \frac{\nabla \cdot \mathbf u^{*}}{\delta t}.
\end{equation}
Now, the correction step (\ref{ISPH corrector}), performed with $p^{n+1}$
obtained from the above relation, yields the divergence-free velocity field.

The common way to solve the Poisson equation in the SPH approach is to compute it directly on particles (irregular grid), cf.~Sect.~\ref{sec:particle poisson solver}. However, there exists another, less useful (specially for free-surface flows), but much more efficient treatment, called here the Grid-Projected Poisson Solver (GPPS). It consists in projecting the r.h.s.\ of (\ref{Poisson}) on an auxiliary grid and then to
crunch it with some commonly known (from Eulerian approaches) solvers. This is a standard technique in various particle methods (more precisely, particle-mesh methods) such as Lagrangian PDF or Vortex-in-Cell.

\subsection{Truly incompressible SPH with the particle Poisson solver}
\label{sec:particle poisson solver}

Employing together the SPH divergence and gradient operators, it is
straightforward to obtain the direct SPH representation of the Laplace operator on the l.h.s.\ of (\ref{Poisson}).
However, Cummins \& Rudman \cite{Cummins & Rudman 1999}
performed a simple one-dimensional hydrostatic test to show that such an approach produces a distinct pressure decoupling pattern.
To avoid this problem, it is common to utilize the approximate Laplacian
operator with the similar form as the viscous N-S term (\ref{SPH NS viscous term})
\begin{equation} \label{PPS pressure 1}
\sph{\nabla \cdot \frac{1}{\varrho} \nabla p}_a \approx \sum_{b} \frac{m_b}{\varrho_a} \frac{4}{\varrho_a + \varrho_b} \frac{p_{ab} \mathbf r_{ab} \cdot \nabla_a W_{ab}(h)}{r_{ab}^2 + \eta^2},
\end{equation}
where $p_{ab} = p_a - p_b$.
The pressure gradient is here computed using a finite-difference approximation.
In this concept the Poisson equation (\ref{Poisson}) is solved on the irregular grid of Lagrangian
points (particles); therefore, this variant of ISPH will be called here the Particle Poisson Solver (PPS).

However, the main disadvantage of the PPS (Sect.~\ref{sec:lid-driven cavity})
is its inefficiency. Dealing with the Poisson equation using Eq.~(\ref{PPS
pressure 1}) consists in solving a linear equation system with a sparse
irregular coefficient matrix. It requires much more CPU time and memory than
the Poisson solvers performed on a regular grid.

\section{Time step criteria}
\label{sec:time step criteria}
In the way to assure the stability of the SPH scheme, several time step criteria must be satisfied \cite{Cummins & Rudman 1999} \cite{Morris et al. 1997} \cite{Cleary & Monaghan 1999} \cite{Monaghan 1992}. In the case of the ISPH approach, the CFL time step condition is
\begin{equation}
\delta t \le 0.25\frac{h}{|\mathbf u|_{max}},
\end{equation}
where $|\mathbf u|_{max}$ is maximal velocity in the flow.
In the WCSPH approach, due to the utilization of the equation of state
(\ref{equation of state}), the CFL time step condition is
\begin{equation}
\delta t \le 0.25\frac{h}{c + |\mathbf u|_{max}}.
\end{equation}
Since we demand density fluctuations to be lower than $1\%$, we have chosen $c
\ge 10 |\mathbf u|_{max}$ (Sect.~\ref{sec:weakly compressible sph}). Therefore,
when the flow is not dominated by viscous or external forces, WCSPH is
computationally less efficient than the ISPH approaches. In the case of explicit schemes for 
viscous flows, another stability criterium is
\begin{equation}
\delta t \le 0.125\frac{h^2}{\nu}. 
\end{equation}
Additional condition must be satisfied due to the magnitude of particle accelerations $\mathbf f$
\begin{equation}
\delta t \le 0.25\min_a \left(\frac{h}{|\mathbf f_a|}\right)^{\frac{1}{2}}.
\end{equation}

\section{Boundary conditions}
\label{sec:boundary conditions}
The proper implementation of the boundaries is one of the common topics in the SPH developments during recent years. Early stage applications of
WCSPH involved high Reynolds number simulations with free-slip boundaries,
performed using one layer of boundary particles placed at the wall. The layer
exerted a strong repulsive force to prevent penetrating solid surfaces
\cite{Monaghan 1989}. Since the number of interacting particles near the walls
is decreased, the accuracy of numerical scheme degrades. Another treatment was
proposed by Campbell \cite{Campbell 1989} where the boundary condition was included
in (\ref{nabla integral interpolant after integration by parts}) through
the residual boundary term. Today, the most often used boundary conditions are
based on dummy particles \cite{Lee et al. 2008} \cite{Shao & Lo 2003}. They are regularly distributed on the boundaries
and have prescribed velocity during the whole
simulation (no-slip condition). In ISPH, the Poisson equation
(\ref{Poisson}) is solved on these dummy particles as well to repulse the fluid.
To prevent inconsistency between density of inner particles and that of the
wall, an additional set of dummy particle layers is placed outside the domain.
Other popular, virtual-particle based boundary conditions utilize so-called mirror
particles.
These particles are given the prescribed velocity to assure the proper boundary condition.
But, their properties are not integrated in time, unlike those of real SPH particles.
Nowadays, there are two commonly used mirror-particle
approaches. The first, developed by Morris \cite{Morris et al. 1997}, consists
in the combination of dummy and mirror particles. The velocity of inner
particles is suitably projected on fixed boundary particles. Then, the boundary
particles interact with the fluid. The second approach, so-called Multiple
Boundary Tangent method, is similar to the previous one but the procedure
of projecting particle velocities is different \cite{Yildiz et al. 2008}.

\begin{figure}
\centering
\includegraphics[width=0.8\textwidth]{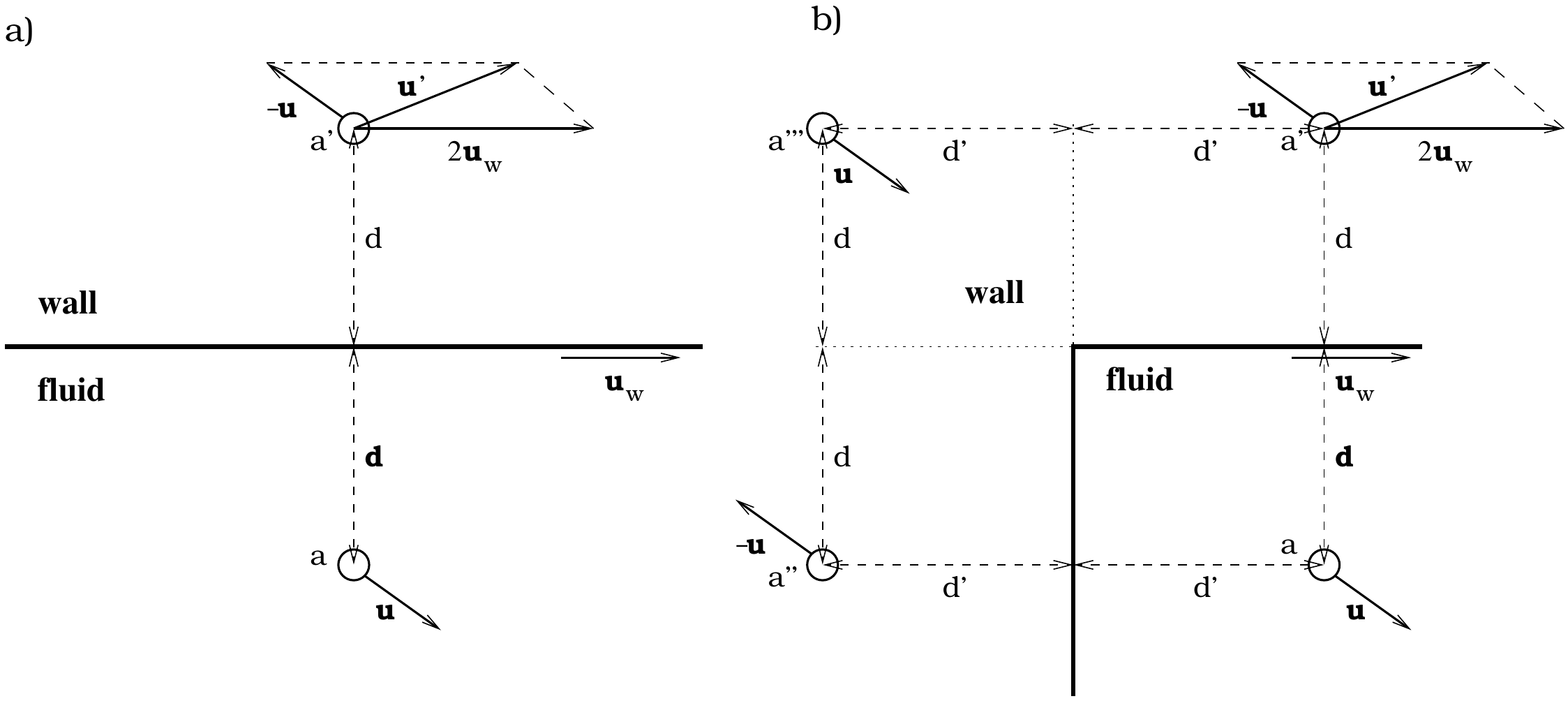}
\caption{The ghost-particle no-slip boundary scheme: a) straight wall, b) inner
corner.}
\label{fig:ghost-particle no-slip boundary scheme}
\end{figure}

The mirror-particle techniques, presented above, have been further developed to another, more natural approach, i.e., the ghost particle method \cite{Cummins & Rudman 1999}. This technique is similar to the Classic Image Problem in electrostatics \cite{Gryffits 1999} \cite{Maxwell 1873}.
To any particle $a$ located at $\mathbf r_a$  near the straight and infinite boundary, we introduce the image $a'$ of this particle located at
\begin{equation} \label{Ghost-particles straight mirror position}
\mathbf r_{a'} = 2\mathbf d + \mathbf r_a
\end{equation}
where $\mathbf d$ is the vector pointing from the particle to the nearest point
at the wall, cf.~Fig.~\ref{fig:ghost-particle no-slip boundary scheme}(a).
Since a chosen kernel is compact, the boundary may be finite. The role of these particles is to assure a high accuracy of the computation (to remedy the lack of particles near the boundaries) and to enforce the boundary condition for the field quantities. Thus, the natural way to obtain the proper implementation of the boundary condition, with the no-slip
condition for velocity field, is to set:
\begin{equation} \label{Ghost-particle straigh no-slip set}
\mathbf u_{a'} = 2\mathbf u_\text{w} - \mathbf u_a, \quad
m_{a'} = m_a, \quad
\varrho_{a'} = \varrho_a,
\end{equation}
where $\mathbf u_\text{w}$ is the velocity of the boundary. To enforce the
proper Neumann boundary condition for the pressure
\begin{equation} \label{boundary condition Neumann pressure}
\frac{\partial p}{\partial n} = 0,
\end{equation}
where $\mathbf n$ is the vector normal to the boundary, 
we extend the set (\ref{Ghost-particle straigh no-slip set}) by an approximation of (\ref{boundary condition Neumann pressure})
\begin{equation}
p_{a'} = p_a.
\end{equation}
Let $\mathbf r_\text{w}$ stand for a position of any point placed at the
boundary. Due to the kernel symmetry we have
\begin{equation} \label{Ghost-particles symmetry across boundary}
W(\mathbf r_\text{w} -\mathbf r_a, h) = W(\mathbf r_\text{w} - \mathbf r_{a'}, h).
\end{equation} 
Now, utilizing the SPH summation interpolant (\ref{SPH interpolant}) for velocity, we write
\begin{equation} \label{velocity on the straight boundary no-slip}
\begin{split}
\sph{\mathbf u}(\mathbf r_\text{w}) &= \sum_a \frac{m_a}{\varrho_a} \mathbf u_a W(\mathbf r_\text{w} - \mathbf r_a, h) + \sum_{a'} \frac{m_{a'}}{\varrho_{a'}} (2\mathbf u_\text{w} - \mathbf u_{a} ) W(\mathbf r_\text{w} - \mathbf r_a', h) = \\
&= 2\mathbf u_\text{w} \sum_{a} \frac{m_a}{\varrho_a} W(\mathbf r_\text{w} -\mathbf r_a, h).
\end{split}
\end{equation}
Since the summation in the above equation is over fluid particles only, assuming
nearly homogeneous distribution of fluid particles, we have
\begin{equation} \label{summing the half of particles near the wall}
\sum_{a} \frac{m_a}{\varrho_a} W(\mathbf r_\text{w} - \mathbf r_a, h) \approx \frac{1}{2}
\end{equation}
and (\ref{velocity on the straight boundary no-slip}) becomes
\begin{equation} \label{proper no-slip condition on straight wall}
\sph{\mathbf u}(\mathbf r_\text{w}) \approx \mathbf u_\text{w}.
\end{equation}
Therefore, (\ref{Ghost-particle straigh no-slip set}) provides the proper formulation of the no-slip condition.

\begin{figure}
\centering
\includegraphics[width=0.6\textwidth]{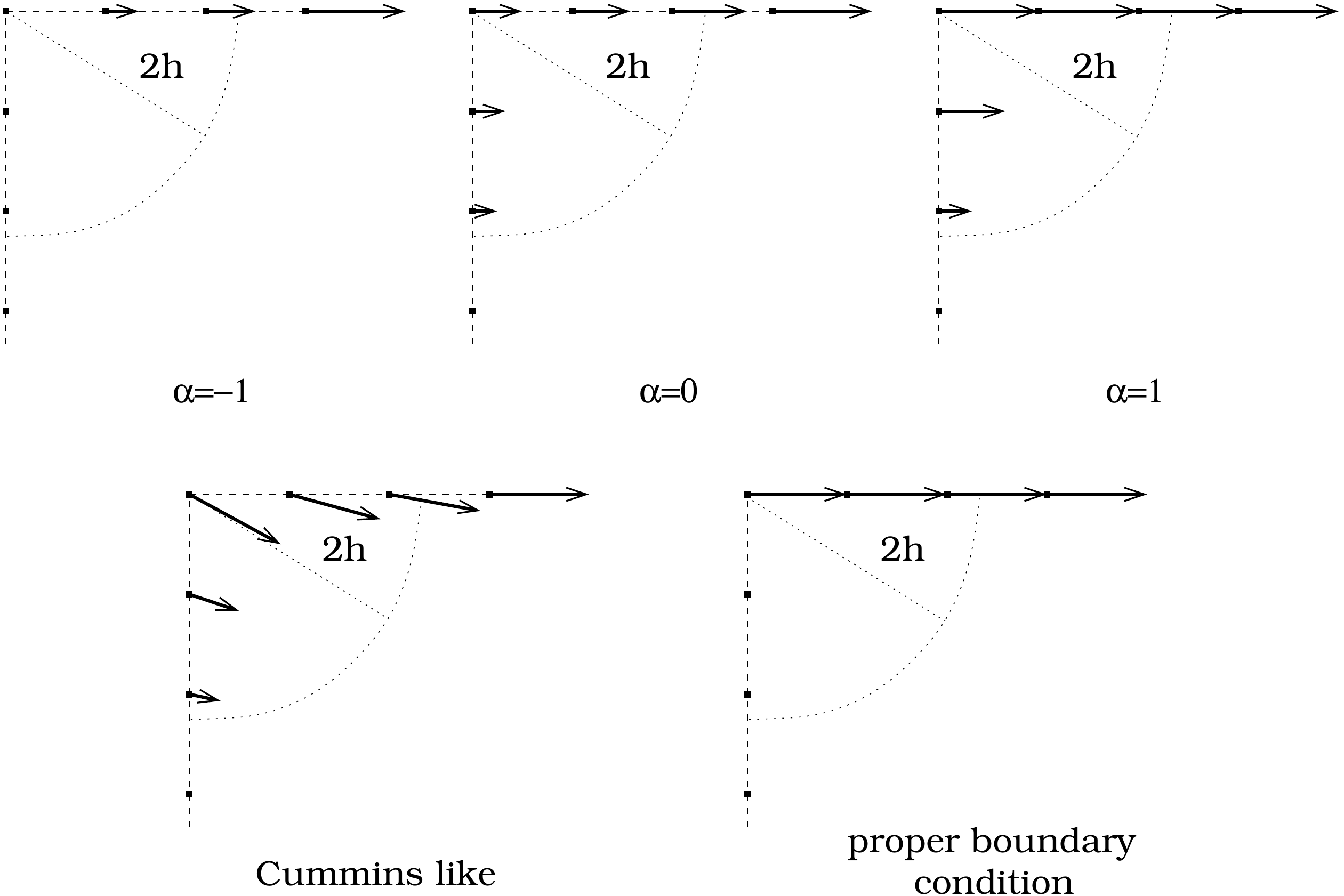}
\caption{The problem with the ghost-particle no-slip boundary condition near the corner: depending on value of the parameter $\alpha$, the velocity boundary condition changes.}
\label{fig:corner problem}
\end{figure}

Another boundary type that can be treated with the ghost-particle approach is an
inner corner. The technique of constructing particle images is presented in
Fig.~\ref{fig:ghost-particle no-slip boundary scheme}(b). In this case we have
to use three mirror particles. It is important to note that the influence range
of this corner is smaller than $2h$. For a larger distance from the corner, the
boundary condition boils down to the previous case (both in vertical and
horizontal directions). For the distances smaller than $2h$, Cummins and Rudman
\cite{Cummins & Rudman 1999} use the third particle placed symmetrically in
respect to the corner point possessing the same density and mass but opposite
velocity to fluid particle. Since the velocity vector computed at the corner is
non-zero in the case of not moving boundaries, this formulation is improper.
Therefore, in the way to find a more accurate approach, we parametrize the
mirror particles' properties as follows (using the notation from
Fig.~\ref{fig:ghost-particle no-slip boundary scheme}):
\begin{equation} \label{Ghost-particle corner no-slip set}
\begin{aligned}
\varrho_a &= \varrho_{a'} = \varrho_{a''} = \varrho_{a'''}, \\
m_a &= m_{a'} = m_{a''} = m_{a'''},
\end{aligned} \quad
\begin{aligned}
\mathbf u_{a'} &= 2 \mathbf u_\text{w} - \mathbf u_{a}, \\
\mathbf u_{a''} &= -\mathbf u_a, \\
\mathbf u_{a'''} &= 2 \alpha \mathbf u_\text{w} + \mathbf u_a,
\end{aligned} \quad
\begin{aligned}
\mathbf r_{a'} &= 2 \mathbf d + \mathbf r_a, \\
\mathbf r_{a''} &= 2 \mathbf d' + \mathbf r_a, \\
\mathbf r_{a'''} &= 2 \mathbf d + 2 \mathbf d' + \mathbf r_a,
\end{aligned}
\end{equation}
where $\alpha$ may change from $-1$ up to $1$. Now, for a point $\mathbf r_w$ at the boundary,
we may carry similar investigation as in (\ref{velocity on the straight boundary
no-slip})
\begin{equation} \label{velocity on the corner no-slip}
\begin{split} 
\sph{\mathbf u}(\mathbf r_\text{w}) &= \sum_a \frac{m_a}{\varrho_a} \mathbf u_a W(\mathbf r_\text{w} - \mathbf r_a, h) + \sum_{a'} \frac{m_{a'}}{\varrho_{a'}} (2\mathbf u_\text{w} - \mathbf u_{a'}) W(\mathbf r_\text{w} - \mathbf r_{a'}, h) \\ 
&+ \sum_{a''} \frac{m_{a''}}{\varrho_{a''}} \mathbf u_{a''} W(\mathbf r_\text{w} - \mathbf r_{a''}, h) + \sum_a \frac{m_{a'''}}{\varrho_{a'''}} (2\mathbf u_\text{w} \alpha + \mathbf u_a''') W(\mathbf r_\text{w} - \mathbf r_{a'''}, h) \\
&= 2\mathbf u_\text{w} \sum_a \frac{m_a}{\varrho_a} \left[ W(\mathbf r_\text{w} - 2\mathbf d + \mathbf r_a, h) + \alpha W(\mathbf r_\text{w} - 2\mathbf d - 2\mathbf d' + \mathbf r_a, h) \right].
\end{split}
\end{equation}
For a point placed exactly at the corner, the above equation reduces to the form
\begin{equation} \label{no-slip condition on corner}
\sph{\mathbf u}(\mathbf r_{\text{corner}}) \approx \frac{1}{2}\mathbf u_\text{w} (1+\alpha).
\end{equation}
Depending on value of the parameter $\alpha$, the velocity of the corner changes from $\mathbf u (\mathbf r_{\text{corner}}) = \mathbf 0$ (for $\alpha = -1$) up to $\mathbf u (\mathbf r_{\text{corner}}) = \mathbf u_\text{w}$ (for $\alpha = 1$). Despite this parametrization, there is no way to assure proper velocity at boundary that is closer than $2h$ from the corner. This issue is presented in Fig.~\ref{fig:corner problem}.
Yet, the simulations of various test problems show that varying parameter
$\alpha$ does not have a significant impact on the global (except near the
corner) velocity or density fields. Therefore, in the present work, we utilize
the no-slip condition with $\alpha=0$.

\begin{figure}
\centering
\includegraphics[width=0.9\textwidth]{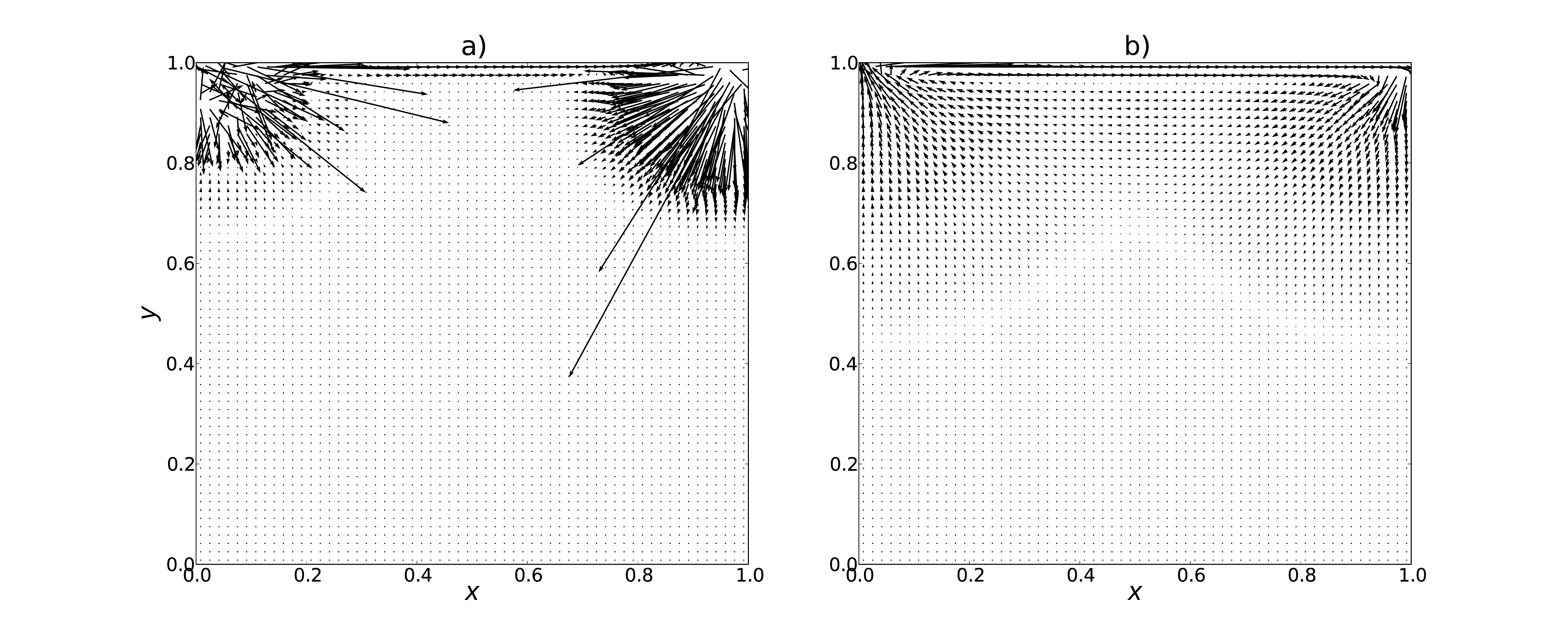}
\caption{The WCSPH results of the lid-driven cavity ($\Reynolds=1000$) at $t=0.03$ with: a) the no-slip and b) the free-slip boundary treatment for velocity divergence computation; employing no-slip condition induces instabilities near the corners.}
\label{fig:instabilities in lid-driven cavity}
\end{figure}

Another problem with the no-slip condition implementation appears during 
computation of density in WCSPH utilizing the continuity equation
(\ref{continuity equation}). Assuming a statistically homogeneous distribution
of particle positions, the velocity component tangential to the
boundary is negligible for computing the divergence. Therefore, along straight
boundaries the no-slip condition is properly stated. The problem appears near
the corners. Utilizing (\ref{SPH nabla interpolant}), we may compute the
divergence of velocity at the boundary near the corner
\begin{equation} \label{div u on the corner no-slip}
\begin{split}
\sph{\nabla \cdot \mathbf u}(\mathbf r_\text{w}) &= \sum_a \frac{m_a}{\varrho_a} \mathbf u_a \cdot \nabla_a W(\mathbf r_\text{w} - \mathbf r_a, h) + \sum_{a'} \frac{m_{a'}}{\varrho_{a'}} \mathbf u_{a'} \cdot \nabla_{a'} W(\mathbf r_\text{w} - \mathbf r_{a'}, h) \\
&+\sum_{a''} \frac{m_{a''}}{\varrho_{a''}} \mathbf u_{a''} \cdot \nabla_{a''} W(\mathbf r_\text{w} - \mathbf r_{a''}, h) +
\sum_{a'''} \frac{m_{a'''}}{\varrho_{a'''}} \mathbf u_{a'''} \cdot \nabla_{a'''} W(\mathbf r_\text{w} - \mathbf r_{a'''}, h).
\end{split}
\end{equation}
For the point $\mathbf r_{c}$ that is placed in the fluid $\varepsilon \rightarrow 0$ away from the corner we have
\begin{equation}\label{corner W symmetries x}
\begin{split}
&\nabla_a W(\mathbf r_c - \mathbf r_a, h) \cdot \mathbf n_x = \nabla_{a'} W(\mathbf r_c - \mathbf r_{a'}, h) \cdot \mathbf n_x 
=\\
 - &\nabla_{a''} W(\mathbf r_c - \mathbf r_{a''}, h) \cdot \mathbf n_x = - \nabla_{a'''} W(\mathbf r_c - \mathbf r_{a'''}, h) \cdot \mathbf n_x
\end{split}
\end{equation}
and
\begin{equation} \label{corner W symmetries y}
\begin{split}
&\nabla_a W(\mathbf r_c - \mathbf r_a, h) \cdot \mathbf n_y =- \nabla_{a'} W(\mathbf r_c - \mathbf r_{a'}, h) \cdot \mathbf n_y =\\
&\nabla_{a''} W(\mathbf r_c - \mathbf r_{a''}, h) \cdot \mathbf n_y = - \nabla_{a'''} W(\mathbf r_c - \mathbf r_{a'''}, h) \cdot \mathbf n_y,
\end{split}
\end{equation}
where $\mathbf n_x$ and $\mathbf n_y$ are unit vectors in $x$ and $y$ directions respectively. Therefore, connecting the above relations with (\ref{div u on the corner no-slip}) and (\ref{Ghost-particle corner no-slip set}) we obtain
\begin{equation} \label{improper div condition on corner}
\sph{\nabla \cdot \mathbf u}(\mathbf r_c) \approx 0.
\end{equation}
Let us consider the situation where fluid particles are driven to the corner. To
prevent penetrating the boundary, there should appear a repulsive force near
this corner. In WCSPH this is done by a local density increase.
However, since the divergence of velocity is always close to zero near the
corner, according to the continuity equation (\ref{continuity equation}),
the density does not change. This may induce a growth of velocity instabilities near the
corners (see Fig.~\ref{fig:instabilities in lid-driven cavity}).
Therefore, only for computing the divergence of velocity, we suggest to use
the free-slip condition.
\begin{figure}
\centering
\includegraphics[width=0.8\textwidth]{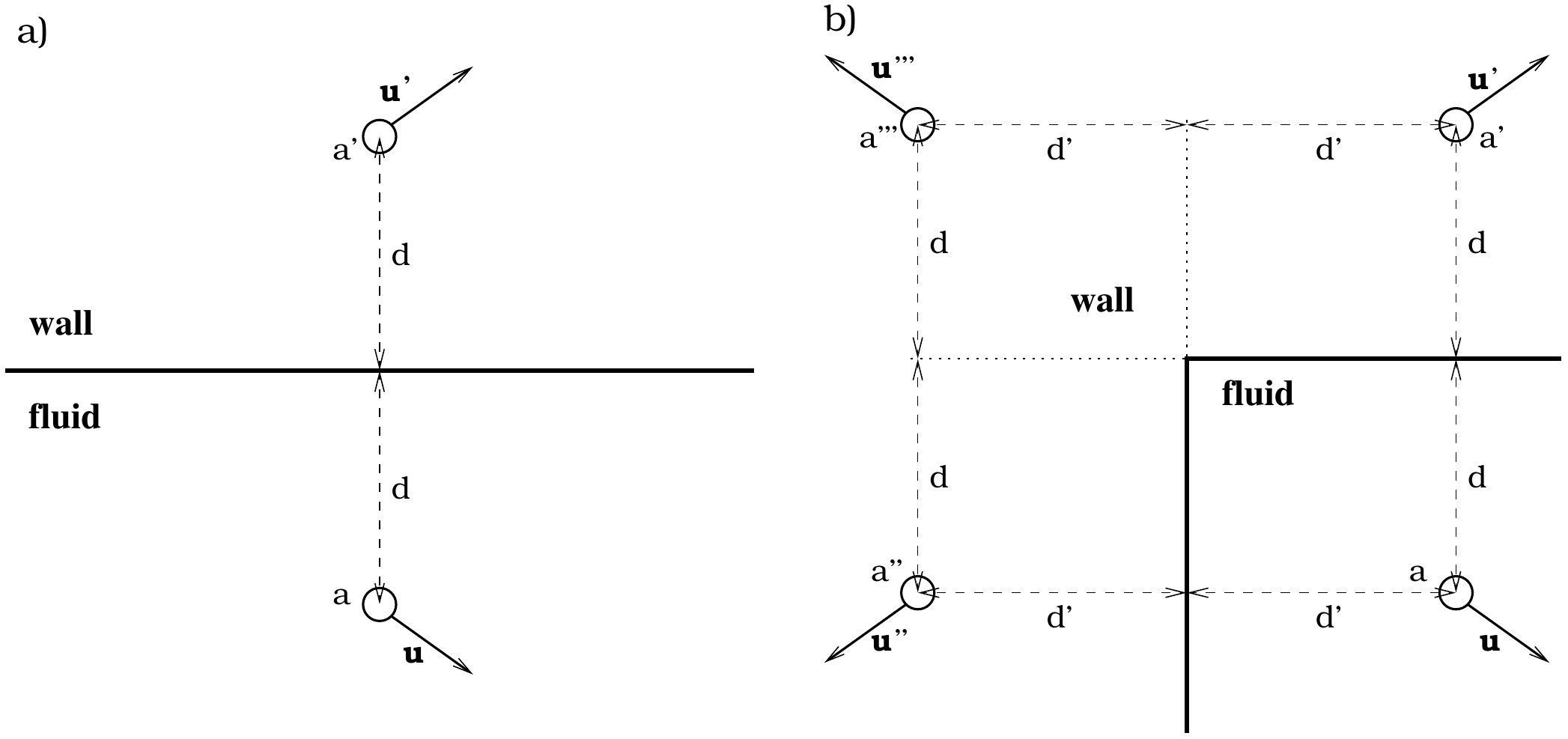}
\caption{The ghost-particle free-slip boundary treatment for: a) the straight wall and b) the corner.}
\label{fig:ghost-particle free-slip boundary scheme}
\end{figure}
For a straight boundary, cf.\ Fig.~\ref{fig:ghost-particle free-slip
boundary scheme}, the free-slip mirror particle is placed
at $\mathbf r_{a'} = 2\mathbf d + \mathbf r_a$. It has the same mass and
density, its velocity component normal to the boundary is opposite, while
the tangential component is unchanged. At the corner, cf.
Fig.~\ref{fig:ghost-particle free-slip boundary scheme}(b), the mirror particles
carry the following properties:
\begin{equation}\label{velocity on the corner free-slip}
\begin{aligned}
\varrho_a &= \varrho_{a'} = \varrho_{a''} = \varrho_{a'''}, \\
m_a &= m_{a'} = m_{a''} = m_{a'''}, \\
\mathbf r_{a'} &= 2 \mathbf d + \mathbf r_a, \\
\mathbf r_{a''} &= 2 \mathbf d' + \mathbf r_a, \\
\mathbf r_{a'''} &= 2 \mathbf d + 2 \mathbf d' + \mathbf r_a,\\
\end{aligned} \quad
\begin{aligned}
\mathbf u_{a'} \cdot \mathbf n_x &=  \mathbf u_a \cdot \mathbf n_x, \\
\mathbf u_{a''} \cdot \mathbf n_x &= -\mathbf u_a \cdot \mathbf n_x, \\
\mathbf u_{a'''} \cdot \mathbf n_x &= -\mathbf u_a \cdot \mathbf n_x,
\end{aligned} \quad
\begin{aligned}
\mathbf u_{a'} \cdot \mathbf n_y &= -\mathbf u_a \cdot \mathbf n_y,\\
\mathbf u_{a'} \cdot \mathbf n_y &=  \mathbf u_a \cdot \mathbf n_y,\\
\mathbf u_{a'} \cdot \mathbf n_y &= -\mathbf u_a \cdot \mathbf n_y.
\end{aligned}
\end{equation}

Obviously, this kind of boundary treatment presented here is basically limited to flat wall segments only; 
more advanced formulations exist and represent the state of the art for arbitrarily-shaped boundaries, 
e.g., ~\cite{Feldman & Bonet 2007}. However, since the main focus of our paper is on incompressibility treatments, 
we have chosen simple geometry cases to have enough reference data.

\section{Velocity error measurement}
\label{sec:velocity error measurement}

\subsection{Lid-driven cavity}
\label{sec:lid-driven cavity}

The lid-driven cavity is a common test of numerical algorithms for viscous flows.
It involves a fluid at density $\varrho_0$ inside a square ($L\times L$) box where only one
boundary moves with the constant velocity $\mathbf u_w$. 
The geometry is very simple, however there is no analytical solution. In the present work, we computed
the lid-driven cavity flow at $Re=|\mathbf u_w|L/\nu=1000$.
For this value of Reynolds number the flow is still laminar and there is no necessity to use a turbulence model.
All results are suitably non-dimensionalised with $L$, $|\mathbf u_w|$, $\varrho_0$ (especially time is normalized with
the convective time scale $L/|\mathbf u_w|$)
and compared to those from a numerical calculation on a fine grid performed with the Eulerian solver by Ghia
et al. \cite{Ghia et al. 1982}.

\subsubsection{The kernel type influence}
\label{sec:the kernels' influence}
\begin{figure}
(a) WCSPH \\
\includegraphics[width=0.95\textwidth]{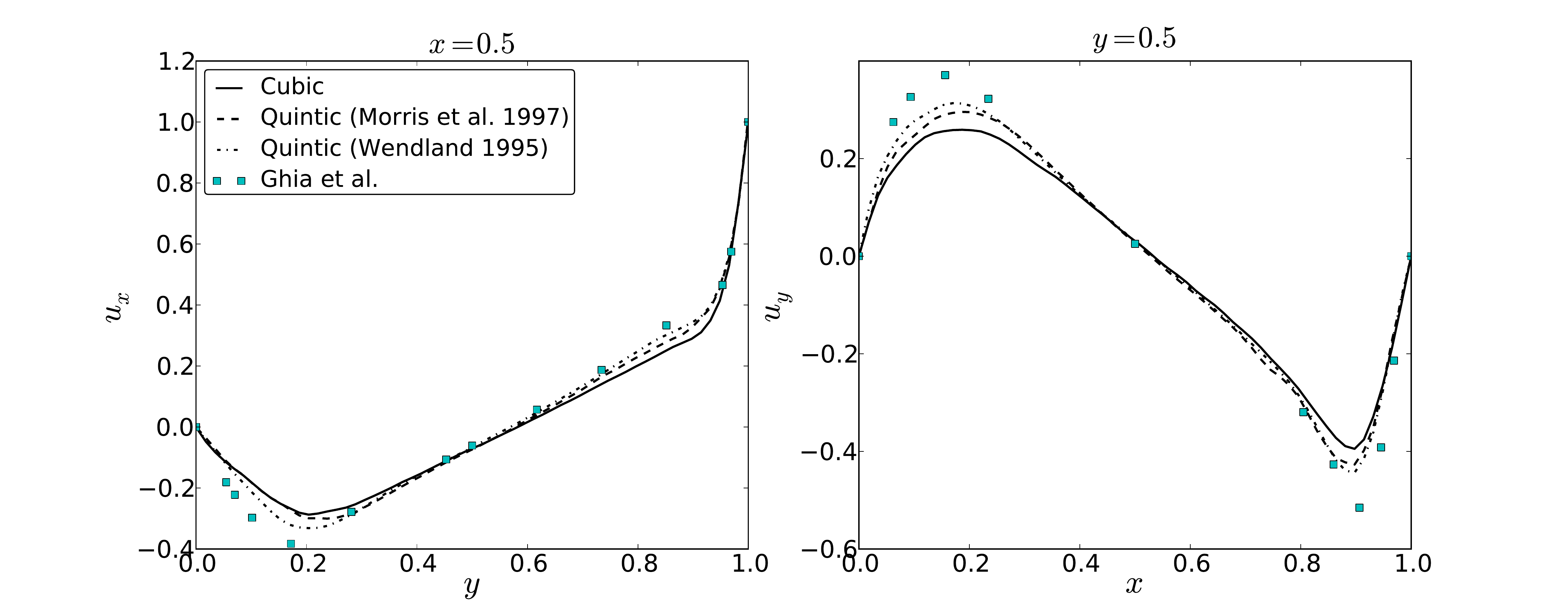} \vspace{10pt} \\ 
(b) ISPH-GPPS \\
\includegraphics[width=0.95\textwidth]{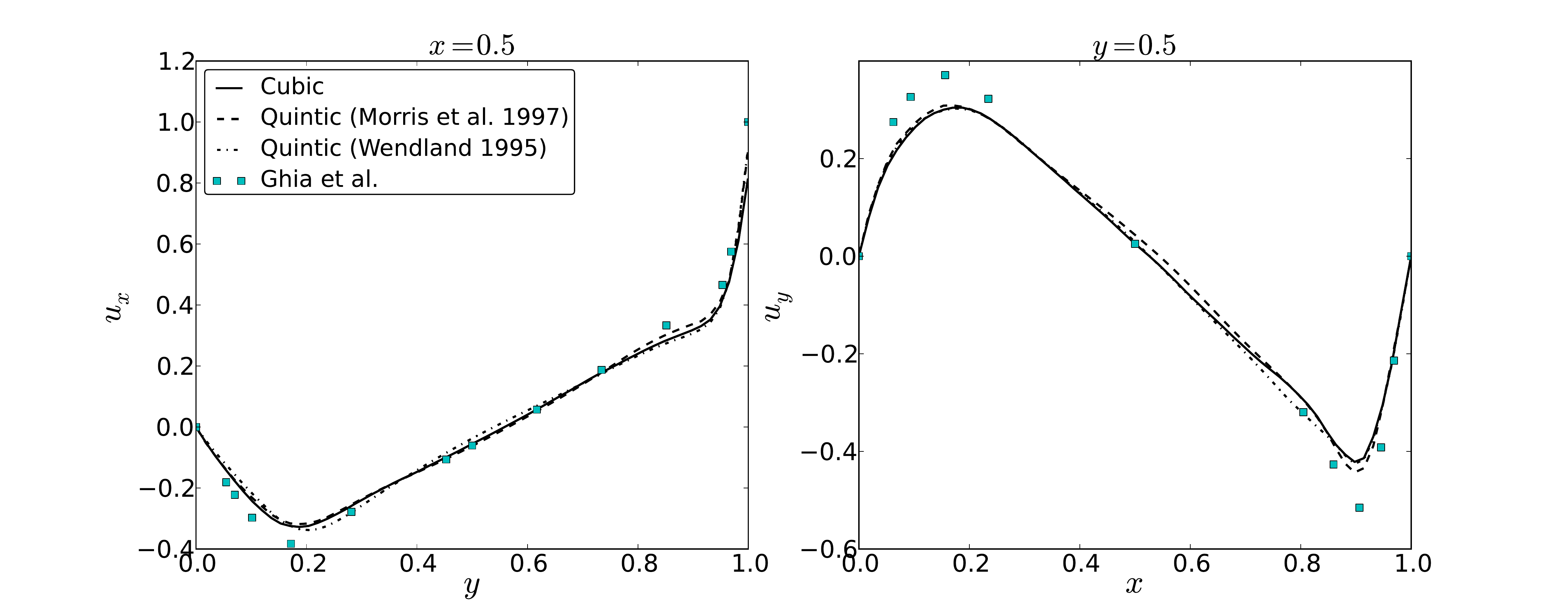} \vspace{10pt} \\
(c) ISPH-PPS \\
\includegraphics[width=0.95\textwidth]{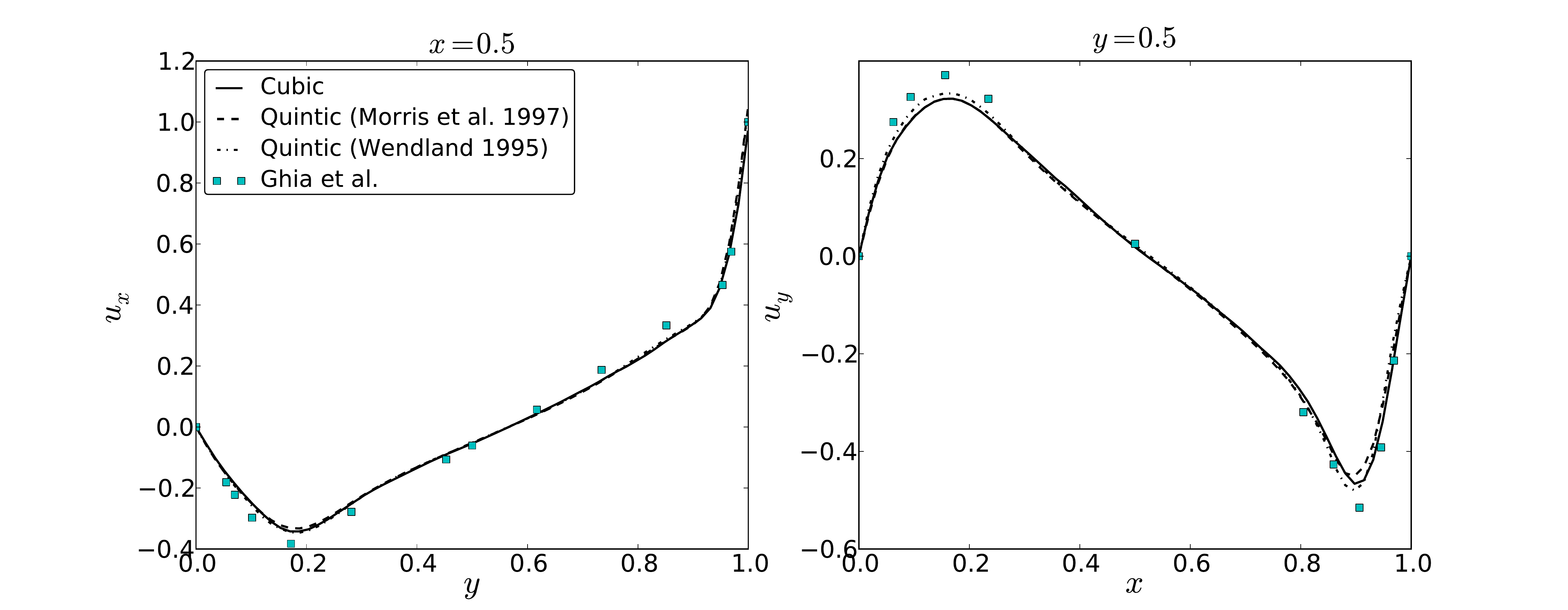}
\caption{The lid-driven cavity steady-state velocity profiles for: (a) WCSPH, (b) ISPH-GPPS and (c) ISPH-PPS against Ghia et al.~\cite{Ghia et al. 1982} results; profiles obtained for different kernels; $N=3600$, $h/\Delta r = 2$.}
\label{fig:lid-sph-kernels}
\end{figure}

\begin{figure}
\centering
\begin{tabular}{ccc}
\includegraphics[width=0.31\textwidth]{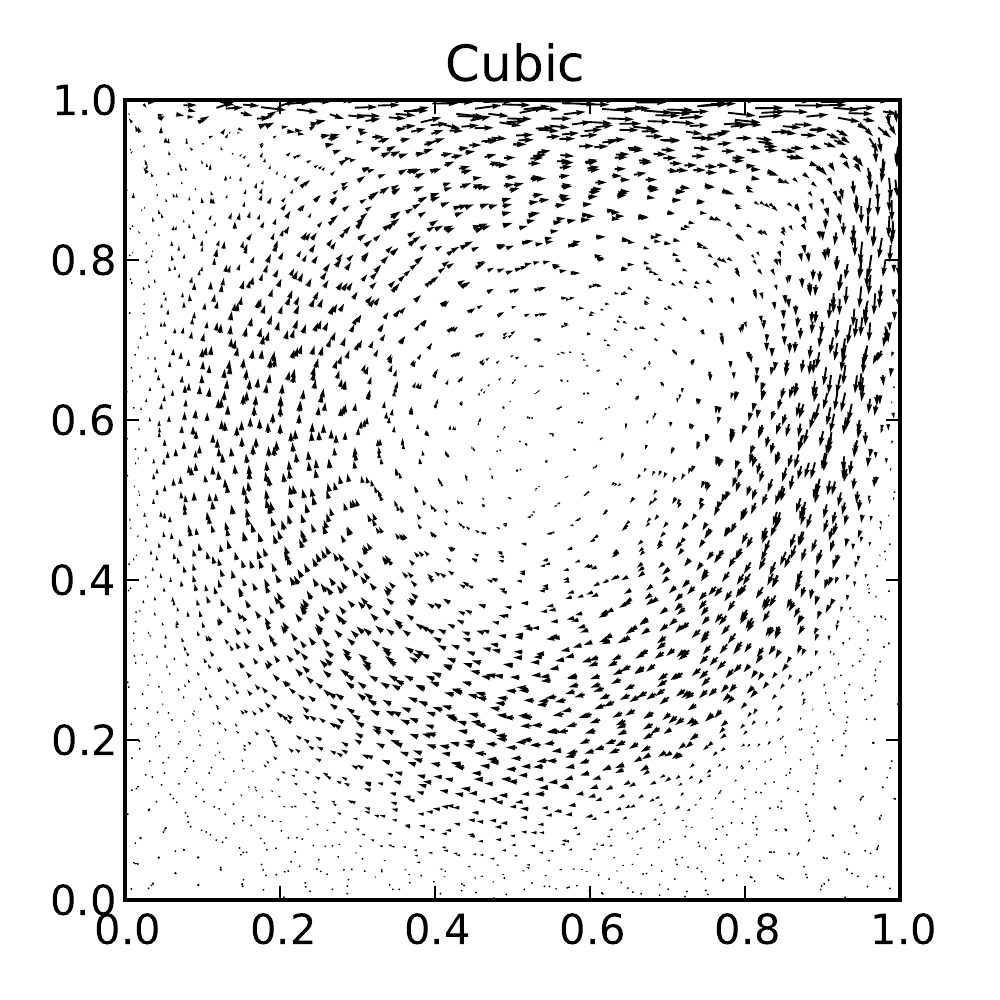} &
\includegraphics[width=0.31\textwidth]{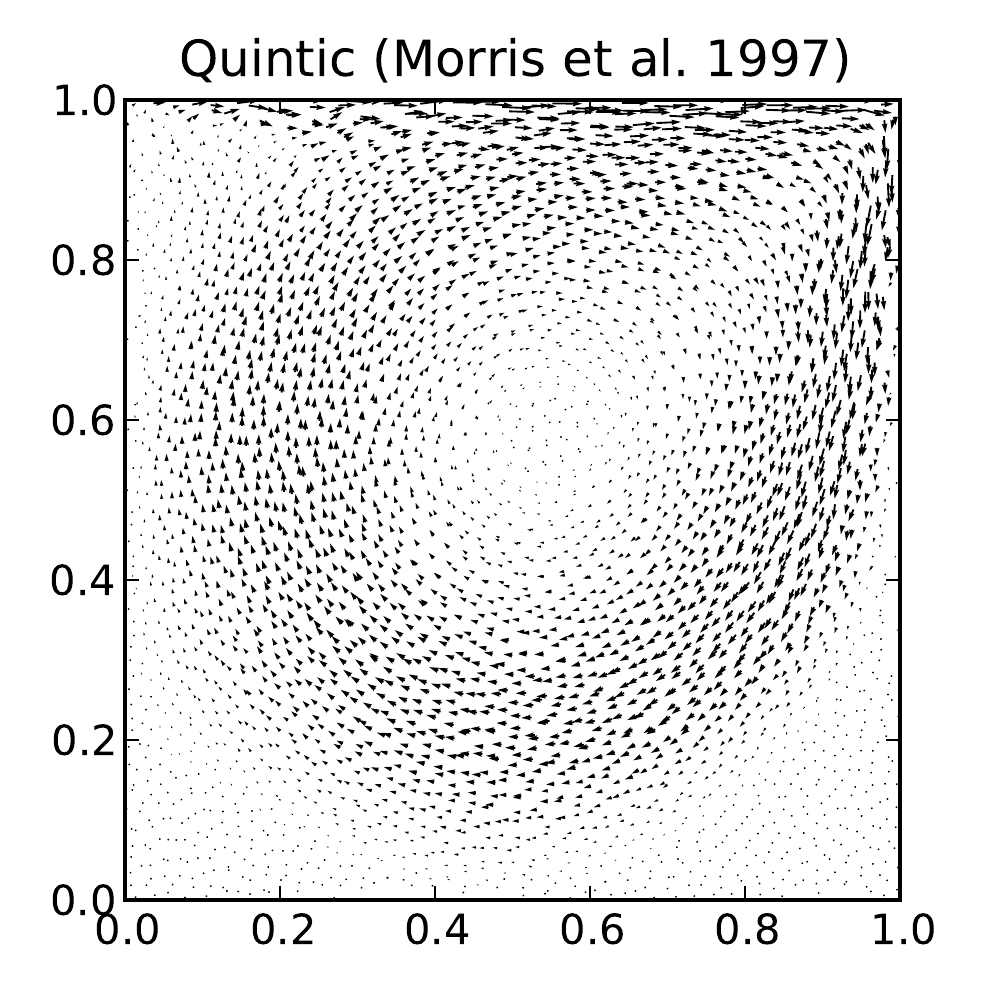} &
\includegraphics[width=0.31\textwidth]{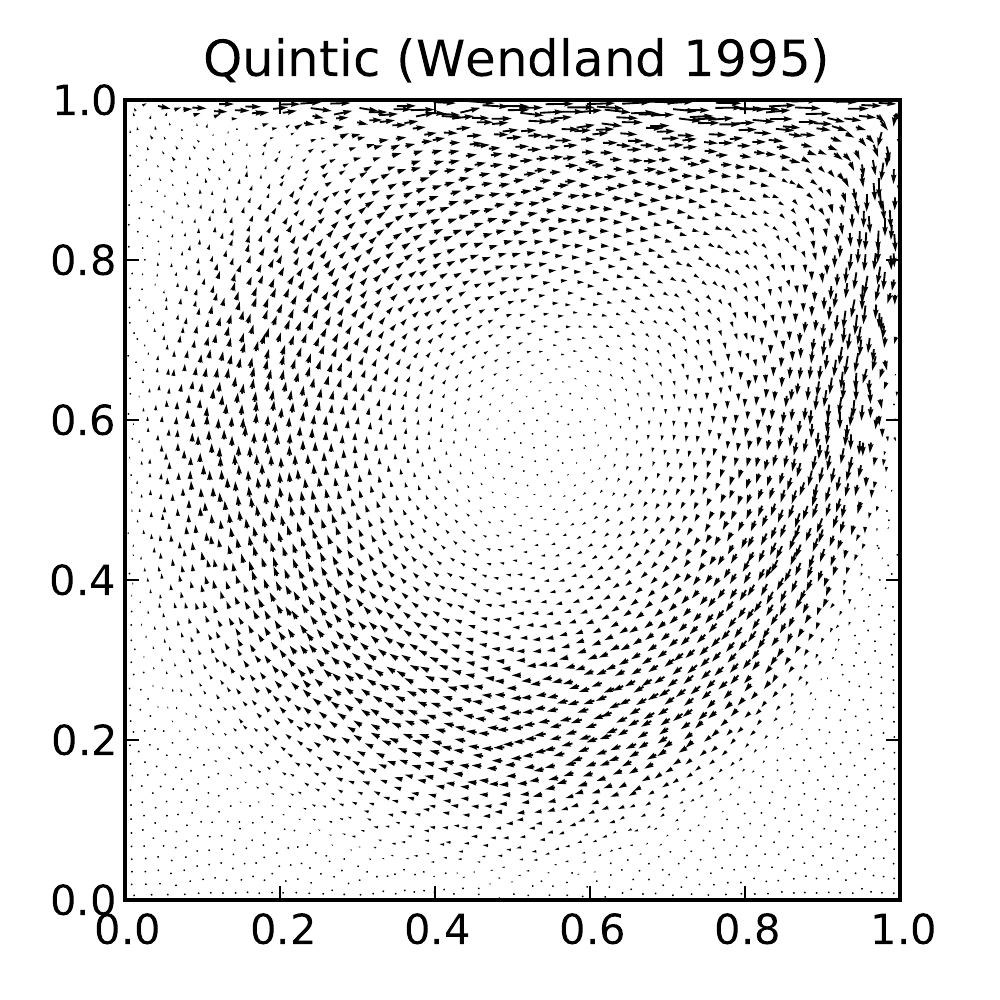}
\end{tabular}
\caption{The lid-driven cavity steady-state solution ($\Reynolds=1000$, $N=3600$) computed with the WCSPH approach and kernels: (\ref{cubic spline kernel}), (\ref{quintic Wendland}), (\ref{quintic Morris}); the particle clustering phenomenon is noticeable with the cubic spline kernel.}
\label{fig:clustering}
\end{figure}

\begin{figure}
\centering
\includegraphics[width=0.70\textwidth]{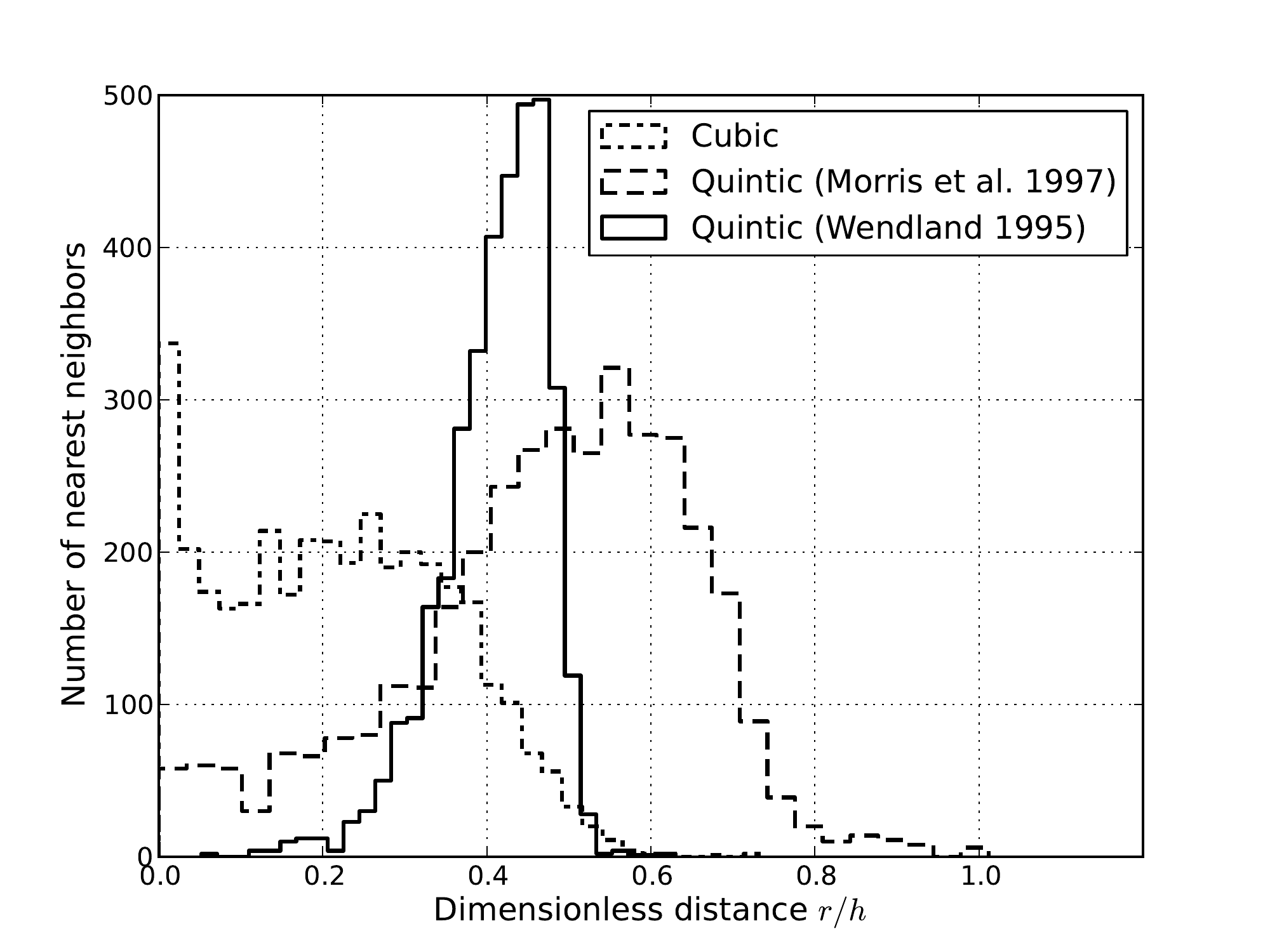}
\caption{Histograms of the distance between the nearest pairs of particles; 
the results obtained for the WCSPH approach and kernels: (\ref{cubic spline kernel}), 
(\ref{quintic Wendland}), (\ref{quintic Morris}).}
\label{fig:histograms}
\end{figure}

One of the important issues that has an impact on the SPH solutions is
a proper selection of the kernel. To compare the treatments of the
incompressibility constraint, we decided to perform the benchmark simulation of the
lid-driven cavity flow with the three kernels presented in
Sect.~\ref{sec:formulation of the SPH method}. The particles'
number has been chosen with two requirements: $N$ should be large enough to get
solutions comparable to reference data (yet not fully converged); however, it should be sufficiently small
to show kernels' defects. We decided to use $N=3600$ particles in the domain and
$h/\Delta r=2$, cf. Sect.~\ref{sec:the smoothing length h influence}. The
steady-state solution velocity profiles are presented in
Fig.~\ref{fig:lid-sph-kernels}. The results
performed with the cubic spline kernel are the most inconsistent with the
reference data. This discrepancy is caused by the particles' clustering
phenomenon. Figs.~\ref{fig:clustering} and \ref{fig:histograms} respectively present particles' spatial distribution
and histograms of the distance between the nearest pairs of particles for all mentioned kernels.
As it transpires from the histogram for the cubic spline kernel, in this case there
are many particles that move joined in groups, cf.\ also details 
in Fig.~\ref{fig:clustering}; as a consequence, the accuracy of the scheme 
is radically decreased. 
As far as the quintic kernels are concerned, both perform
similarly to each other. However, the kernel proposed by Morris et al.~(\ref{quintic
Morris}) shows a more pronounced tendency to clustering 
(Fig.~\ref{fig:histograms}); moreover, it is not zero up to $|\mathbf r|/h = 3$ (as contrasted to $|\mathbf r|/h = 2$ for the
other kernels), so it involves many more particles. Therefore, the
computational time is increased about 12\% comparing to the Wendland kernel. 
Summarizing, due to a good agreement with the reference data, not noticeable clustering phenomenon, and
finally the efficiency, for the further analysis we decided to use the quintic
Wendland kernel. An interesting work about its behavior in SPH has recently been performed by Robinson~\cite{Robinson 2009} (Ch.\ 7).

\subsubsection{The kernel size influence}
\label{sec:the smoothing length h influence}
\begin{figure}
(a) WCSPH \\
\includegraphics[width=0.95\textwidth]{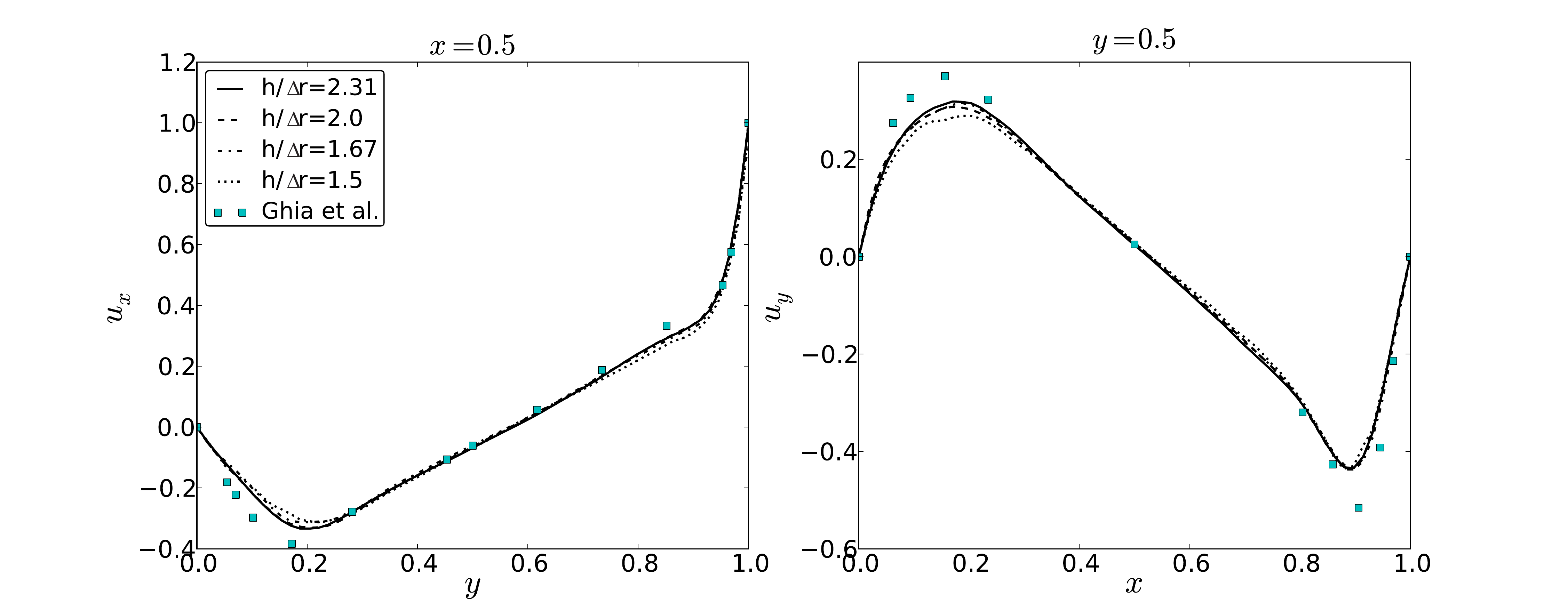} \vspace{10pt} \\ 
(b) ISPH-GPPS \\
\includegraphics[width=0.95\textwidth]{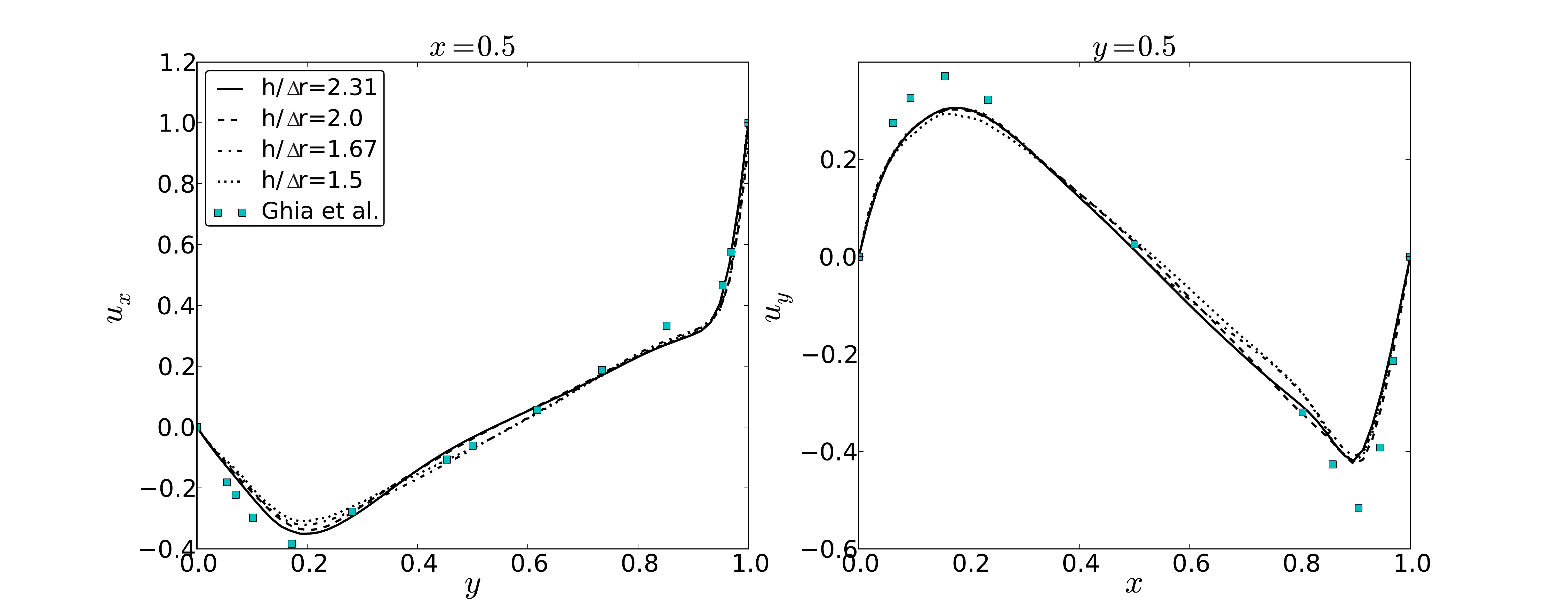} \vspace{10pt} \\
(c) ISPH-PPS \\
\includegraphics[width=0.95\textwidth]{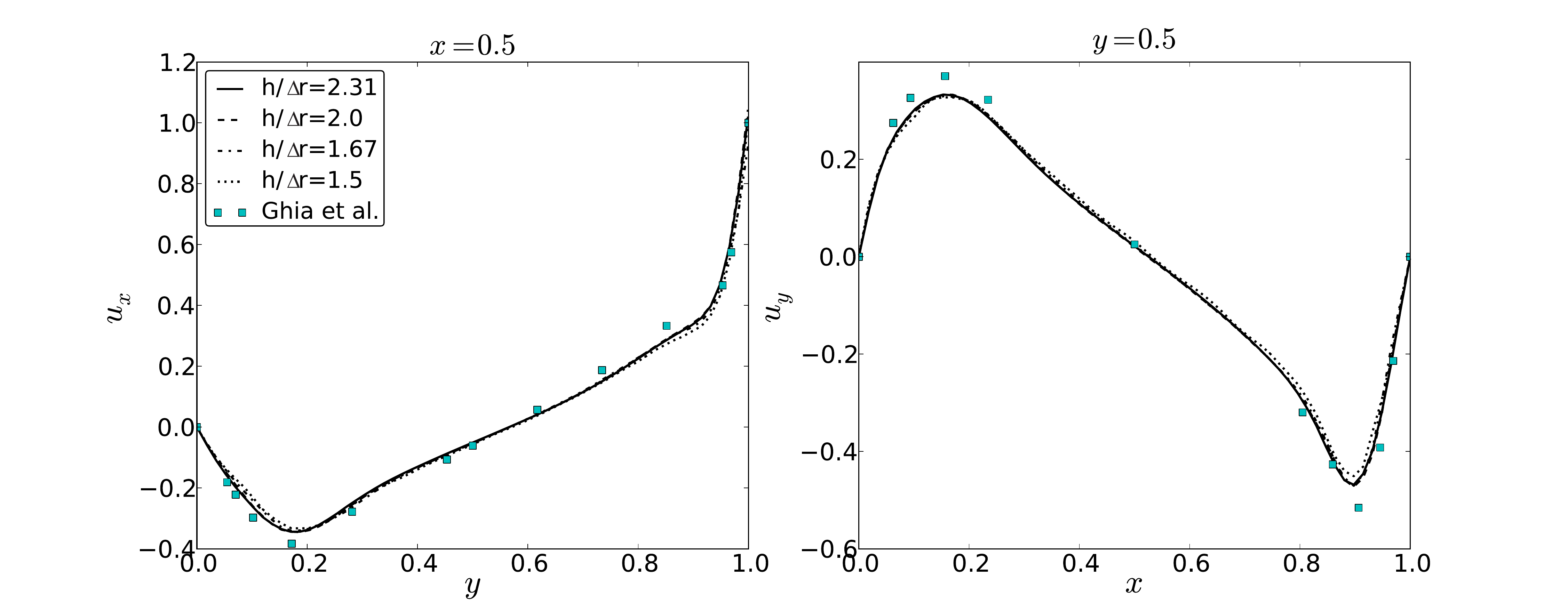}
\caption{The lid-driven cavity steady-state velocity profiles for: (a) WCSPH, (b) ISPH-GPPS and (c) ISPH-PPS against Ghia et al.~\cite{Ghia et al. 1982} results; profiles obtained with different $h/\Delta r$ values; results obtained using the Wendland kernel~\cite{Wendland 1995} and $N=3600$ particles in domain.}
\label{fig:lid-sph-h}
\end{figure}

Apart from  the initial inter-particle distance $\Delta r$ (related to the
number of particles in the domain), in the SPH approach there exists another
characteristic length - the kernel parameter $h$. Examining the impact on the
solutions, we decided to compute the lid-driven cavity problem ($Re=1000$) for
chosen $N=3600$, the Wendland kernel and different $h/\Delta r$ values.
Figure~\ref{fig:lid-sph-h} presents the velocity profiles for $h/\Delta r=2.31$, $2.0$, $1.67$ and $1.5$
computed with all the SPH incompressibility variants. In comparison to the Ghia
et al. reference data, there is no significant effect of the kernel size on the velocity field.
However, since the parameter $h/\Delta r$
determines the number of particles under the kernel hat, the number of interaction between particles 
(and the computational effort) grow like $\sim (h/\Delta
r)^D$, where $D$ is the space dimension. Weighting between the computational
times and the accuracy, for all simulations presented henceforth, we
decided to use $h/\Delta r=2.0$.

\subsubsection{The particle number influence}
\label{sec:the particles' number N influence}
\begin{figure}
(a) WCSPH \\
\includegraphics[width=0.95\textwidth]{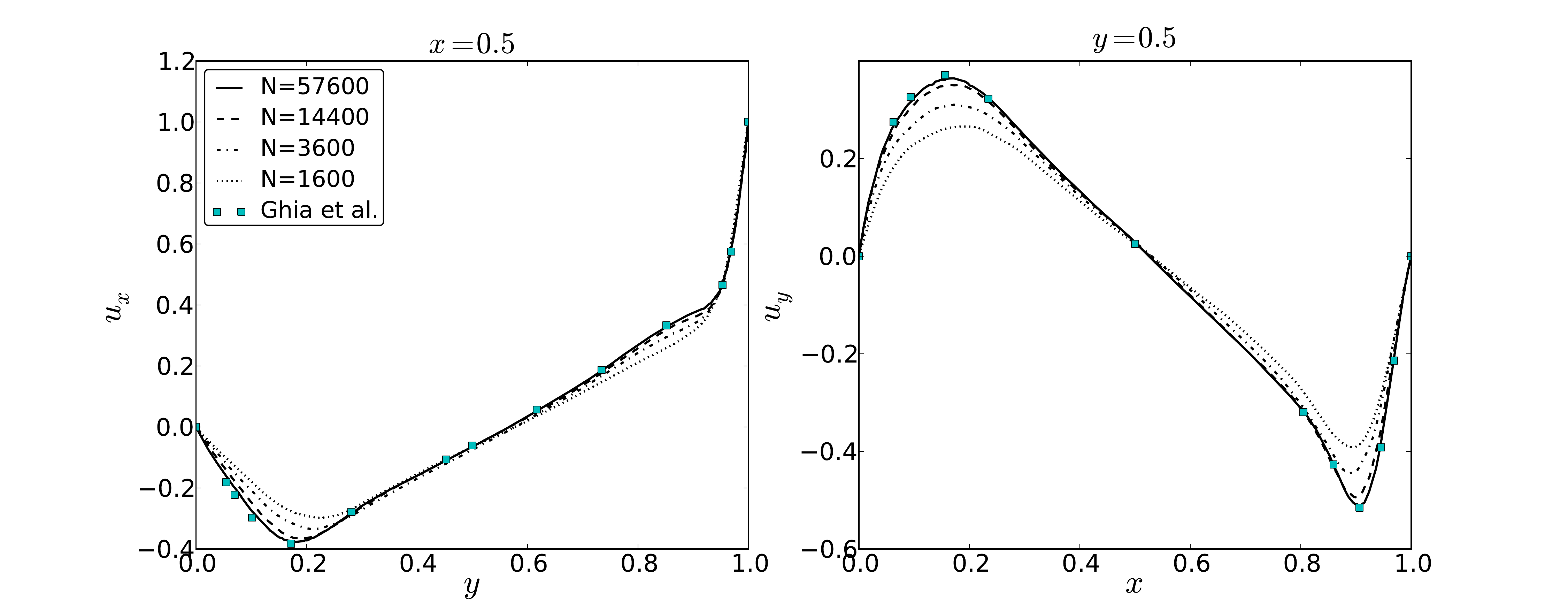} \vspace{10pt} \\ 
(b) ISPH-GPPS \\
\includegraphics[width=0.95\textwidth]{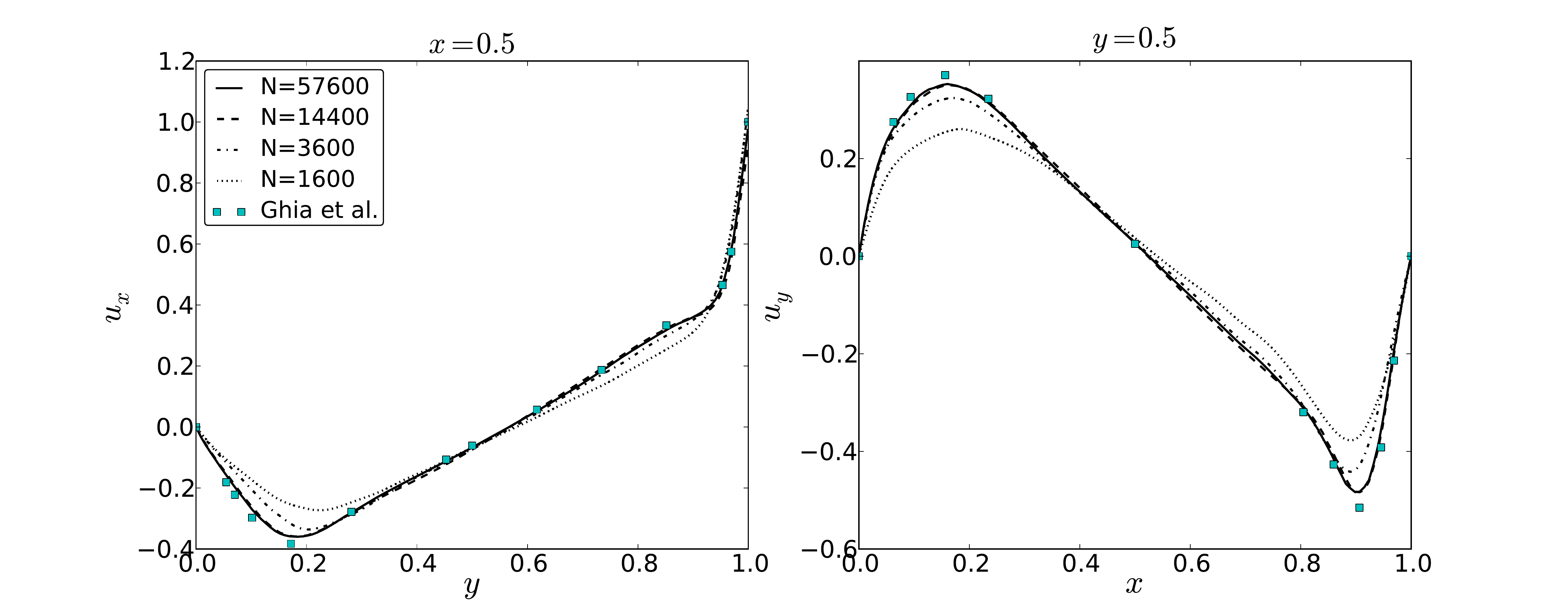} \vspace{10pt} \\
(c) ISPH-PPS \\
\includegraphics[width=0.95\textwidth]{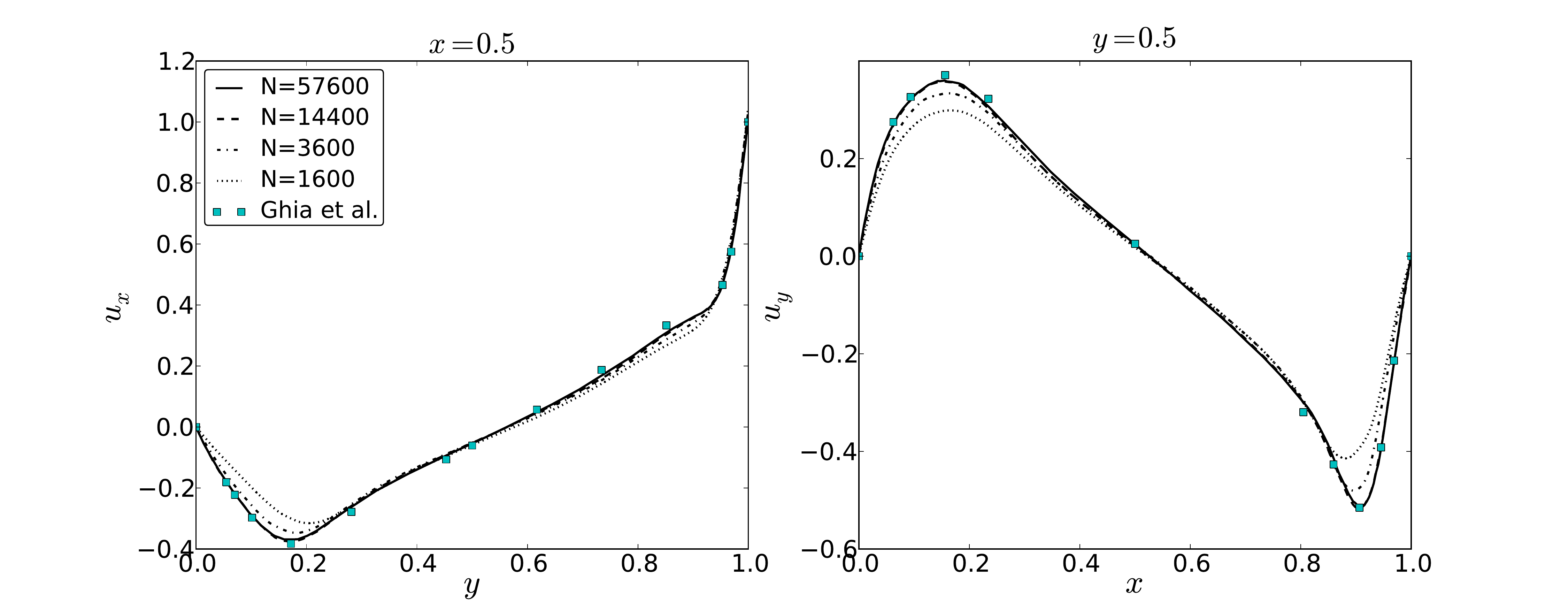}
\caption{The lid-driven cavity steady-state velocity profiles for: (a) WCSPH, (b) ISPH-GPPS and (c) ISPH-PPS  against Ghia et al.~\cite{Ghia et al. 1982} results; results for different number of particles $N$; data obtained using the Wendland kernel~\cite{Wendland 1995} and $h/\Delta r = 2$.}
\label{fig:lid-sph-n}
\end{figure}

Examining the influence of the spatial resolution, the simulations were
performed with a different number of particles in the domain (from $N=1600$ up
to $N=57\,600$). The steady-state velocity profiles are presented in Fig.~\ref{fig:lid-sph-n}.
In all the methods of
incompressibility treatment, the profiles obtained for $N=1600$ did not agree
very well with the Ghia solution \cite{Ghia et al. 1982} obtained on $129
\times 129$ Eulerian mesh. Performing simulations with the higher resolutions,
the accuracy increases so that in the case of $N=14\,400$, independently on the
compressibility treatment, the obtained results are in very good agreement with
the reference simulations.
For $N=57\,600$ the velocity profiles computed using WCSPH and ISPH - PPS
approaches practically overlap with the Ghia's reference data. Unfortunately,
the ISPH - GPPS solution is not so accurate as we expect. This deficiency is
due to a higher numerical diffusion caused by the projection of the
quantities from particles on a regular grid. 

Due to the utilization of the equation of state and the time step constraint in the WCSPH approach, the CPU
time of the lid-driven cavity simulation computed by the ISPH method with PPS is
about $7$ times shorter. On the other hand, performing the ISPH-GPPS simulations, the
computational effort may be reduced about $15$ times. The comparison of CPU
times for all considered SPH schemes is presented in Fig.~\ref{fig:lid-driven
cavity times}. 

\begin{figure}
\centering
\begin{tabular}{ccc}
\includegraphics[width=0.6\textwidth]{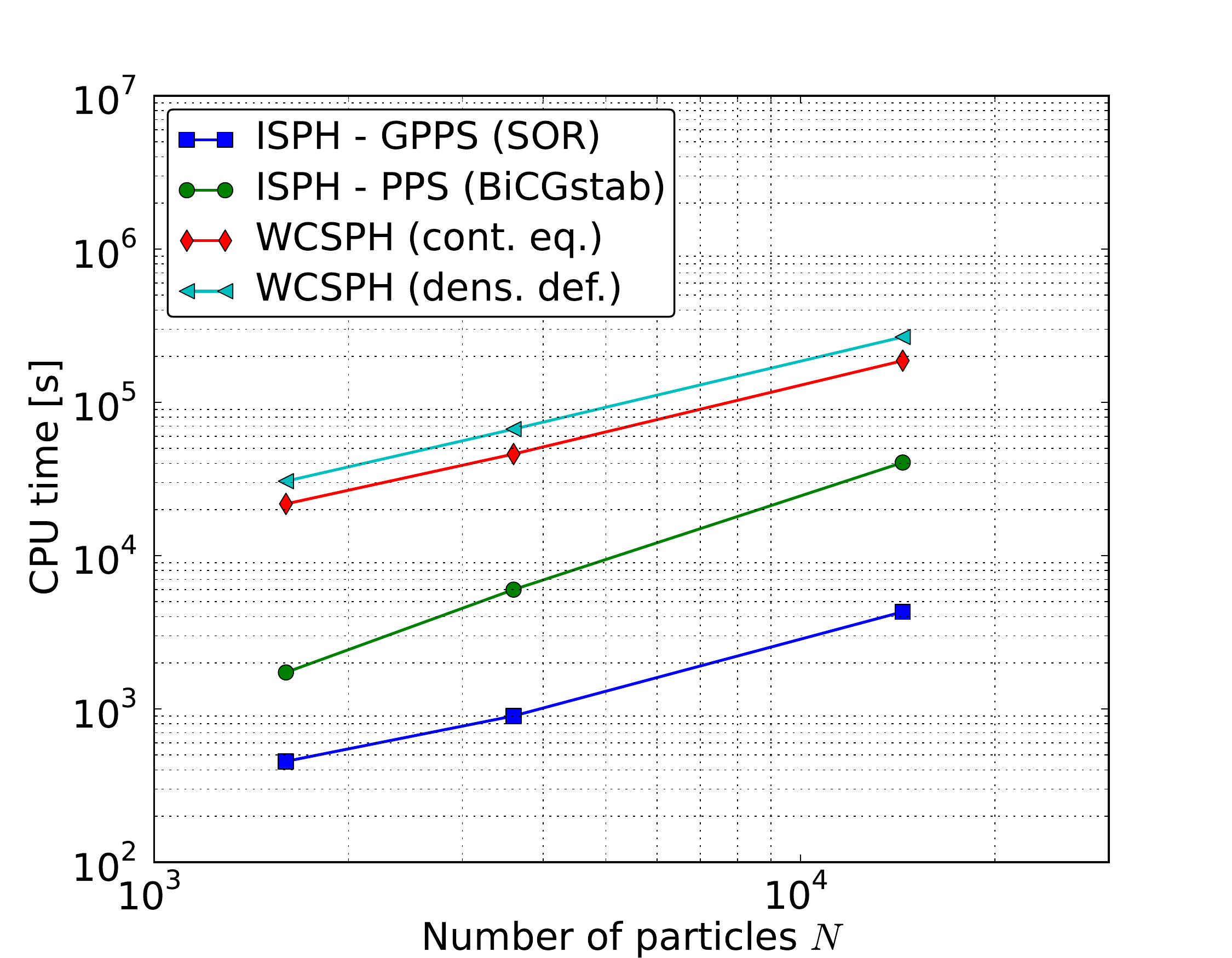} &
\end{tabular}
\caption{The CPU times to obtain the steady-state solution of the lid-driven cavity ($\Reynolds=1000$) using the WCSPH and both ISPH approaches; for the WCSPH method, two continuity equations are compared: Eq.~(\ref{SPH continuity symmetrical}) and density definition (\ref{SPH direct density computation multiphase}).}
\label{fig:lid-driven cavity times}
\end{figure}

\subsection{The Rayleigh-Taylor instability}
\label{sec:the rayleigh-taylor instability}
\begin{figure}
\centering
\begin{tabular}{ccc}
 \includegraphics[width=0.33\textwidth]{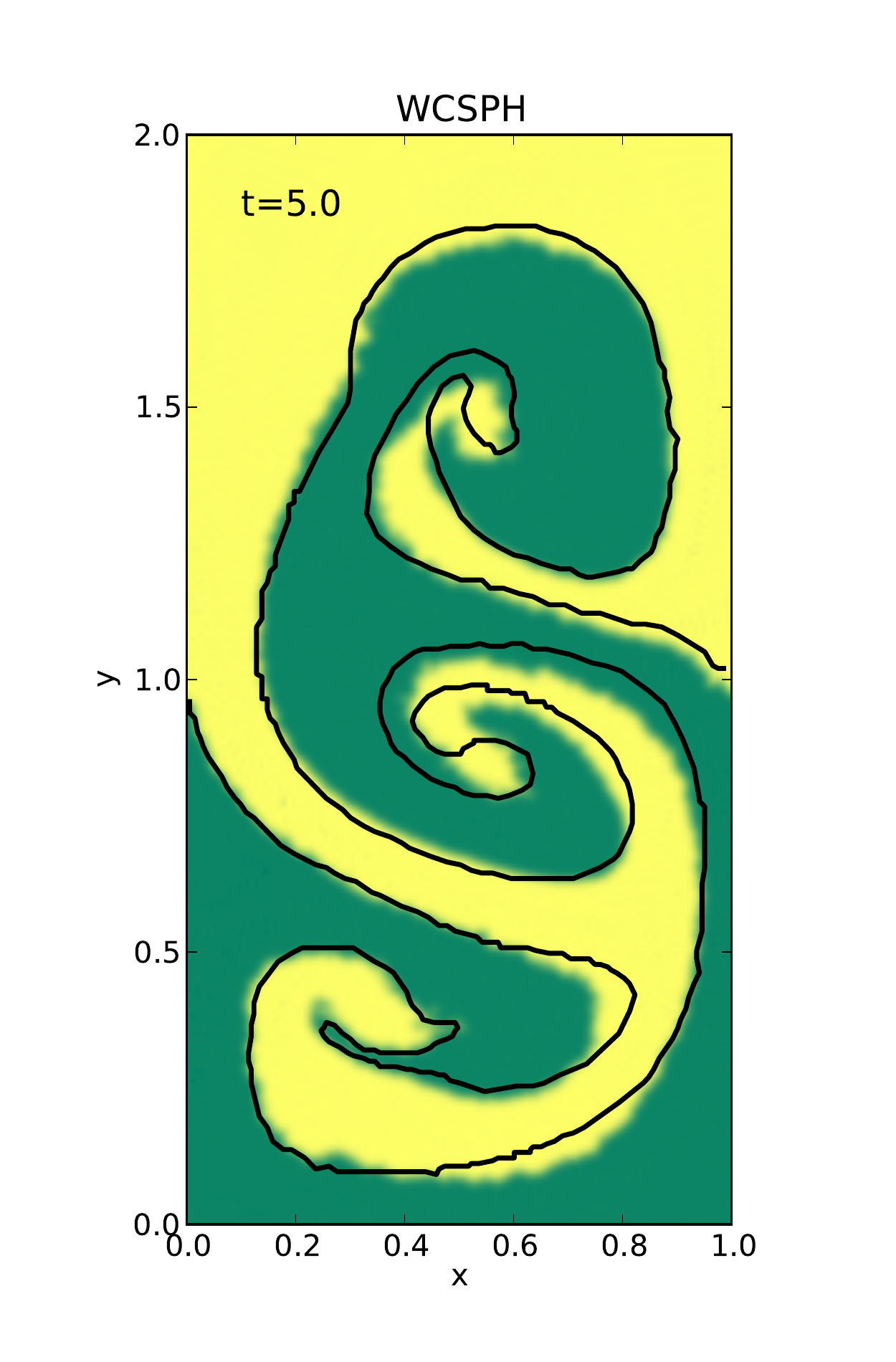}
 \includegraphics[width=0.33\textwidth]{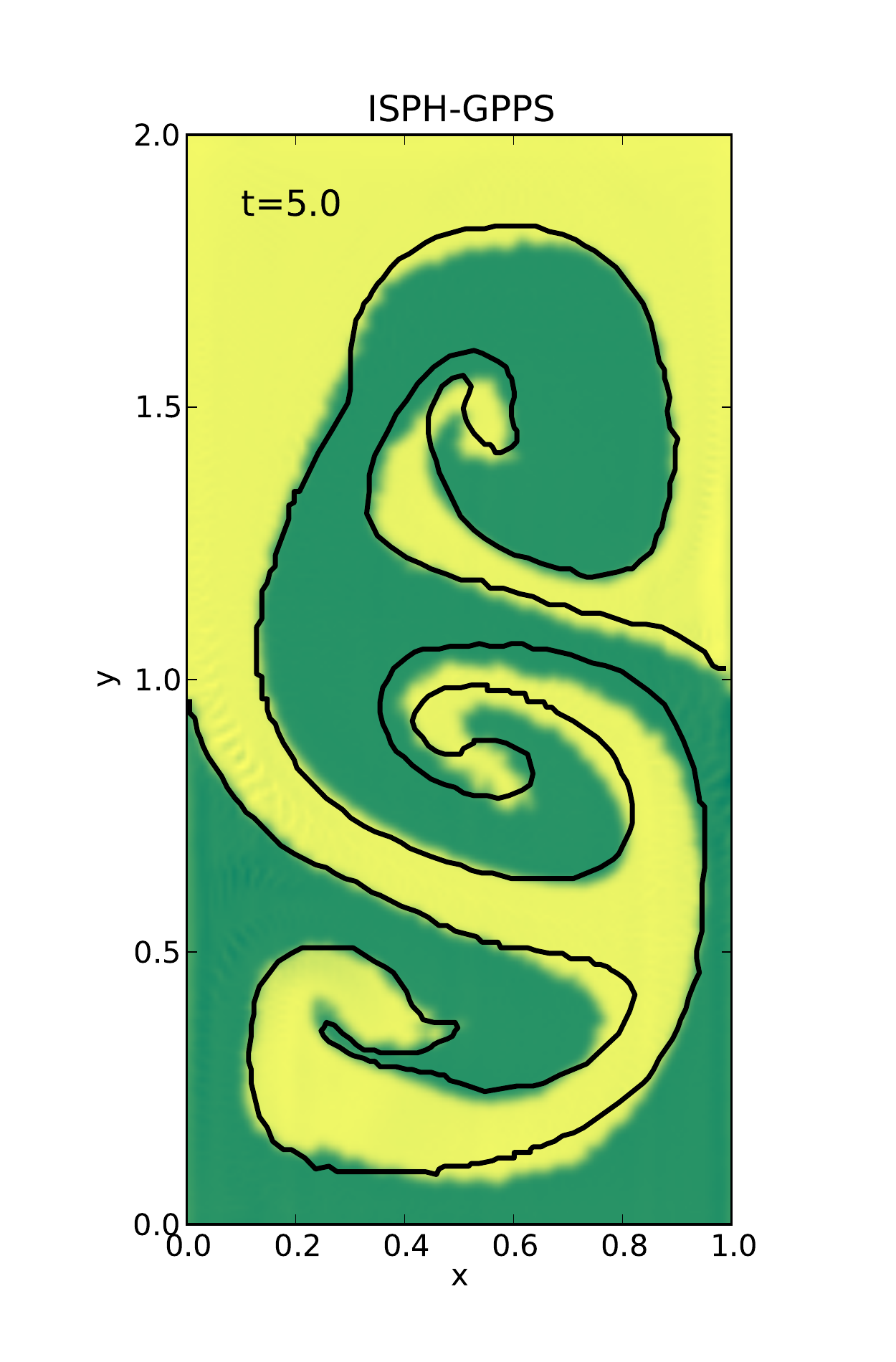}
 \includegraphics[width=0.33\textwidth]{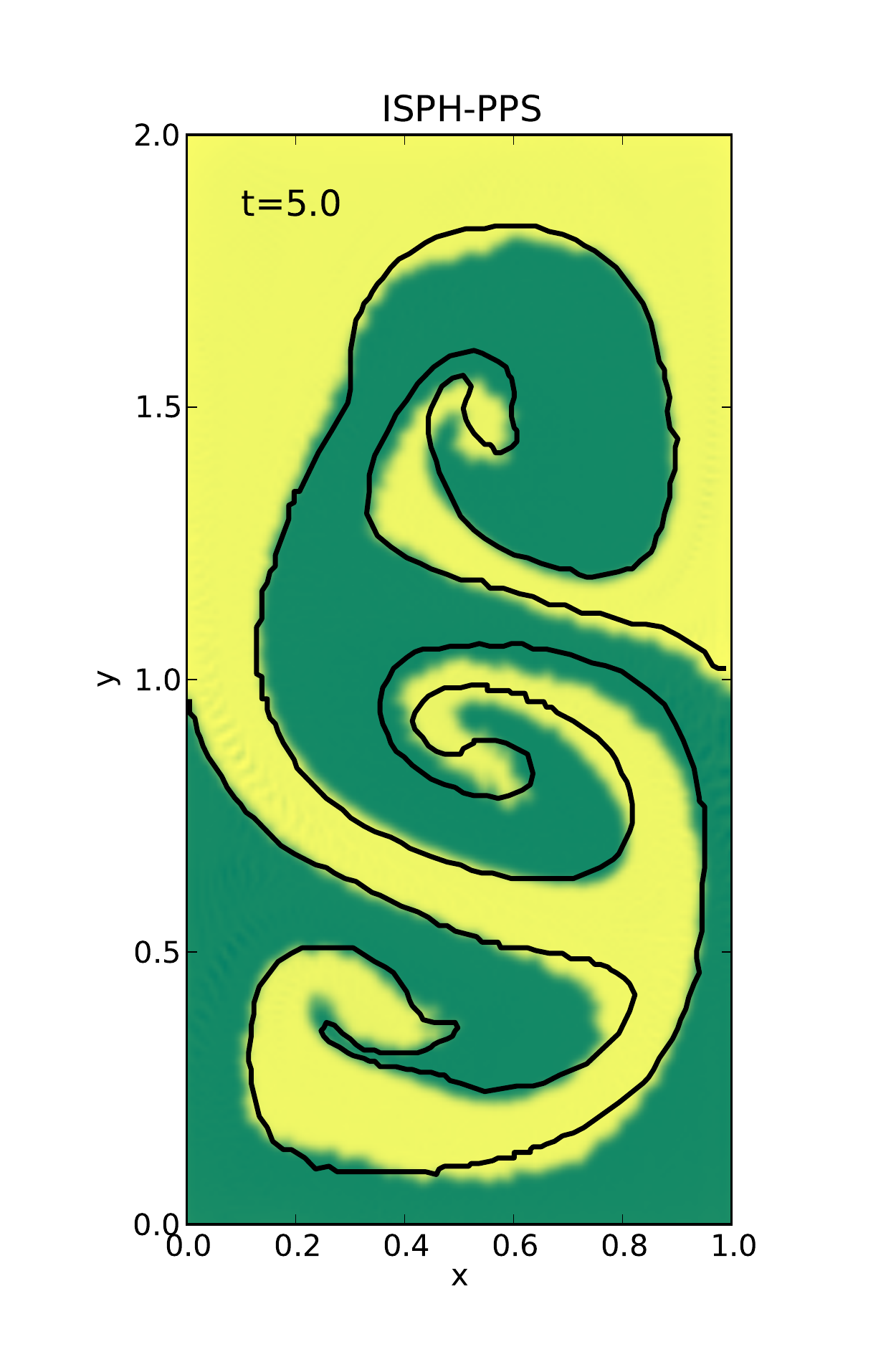}
\end{tabular}
\caption{The Rayleigh-Taylor instability; particle positions at $t=5$ obtained using different incompressibility treatments; solid line: liquid-liquid interface from the reference Level-Set solution \cite{Grenier et al. 2009}.}
\label{fig:rt-all}
\end{figure}

The Rayleigh-Taylor instability is one of generic multi-phase flows, therefore it is commonly utilized as a test problem.
It involves two immiscible fluids enclosed in a rectangular domain of width
$L$ and height $2L$.
Initially, in our case, the phases are separated by the interface located at $y=1-0.15\sin{(2\pi x)}$. The lower component has density $\varrho_L=\varrho_0$, while the upper one $\varrho_U=1.8 \varrho_0$. Since the system is subject to gravity $\mathbf g=(0, -g)$ and the upper phase is heavier, in the absence of the surface tension an instability always arises and vorticity is generated. The Reynolds number may be defined as
\begin{equation}
\Reynolds = \frac{\sqrt{L^3 g}}{\nu} = 420.
\end{equation}
The simulations were performed for $120\times 240$ particles in the domain. 
To compute the hydrostatic force, we use the technique described in~\cite{Szewc et al. 2011},
 where the hydrostatic pressure is computed on a regular mesh and later projected on the particles.
Figure~\ref{fig:rt-all} presents particle positions (directly
showing the location of liquid-liquid interface) computed with different
treatments of the incompressibility condition. All the simulations are compared
to the reference solutions from the Level-Set method ($312\times 624$ cells)
computed by Grenier et al.~\cite{Grenier et al. 2009}. Presented data were
obtained at $t=5$ (normalized with the convective time scale $\sqrt{L/g}$).

Comparing the interface shapes, we find no convincing arguments
to judge which SPH approach is more accurate. However, the comparison of the
computational times shows that the most valuable choice is the ISPH-GPPS. The
ISPH-PPS is slower about $6$ times, while the WCSPH about $13$ times.

\section{Density error measurement}
\label{sec:density error measurement}

In order to measure the density estimation errors, we compute two quantities:
the mean density over the flow domain $\mean \varrho(t)$ and the root mean
square of the density fluctuations $\varrho_{\text{rms}}(t)$. However, since in the
Projection Methods the values of the density carried by particles do not change,
there is the necessity to use formulas which are able to compute $\mean
\varrho(t)$ and $\varrho_{\text{rms}}(t)$ taking into account both the density
values and spatial distribution of the particles. The simplest proposals (expressed
in the SPH formulae) are: for the mean density 
\begin{equation}
\mean \varrho(t) = \frac{1}{N} \sum_a m_a \sum_b W_{ab}(h),
\end{equation}
and for the density fluctuations $\varrho_{\text{rms}}(t)$
\begin{equation}
\varrho_{\text{rms}}(t) = \sqrt{ \frac{1}{N} \sum_{a} \left( m_a \sum_b W_{ab} - \mean \varrho(t) \right)^2}.
\end{equation}

\begin{figure}
\centering
\begin{tabular}{cc}
(a)
\includegraphics[width=0.45\textwidth]{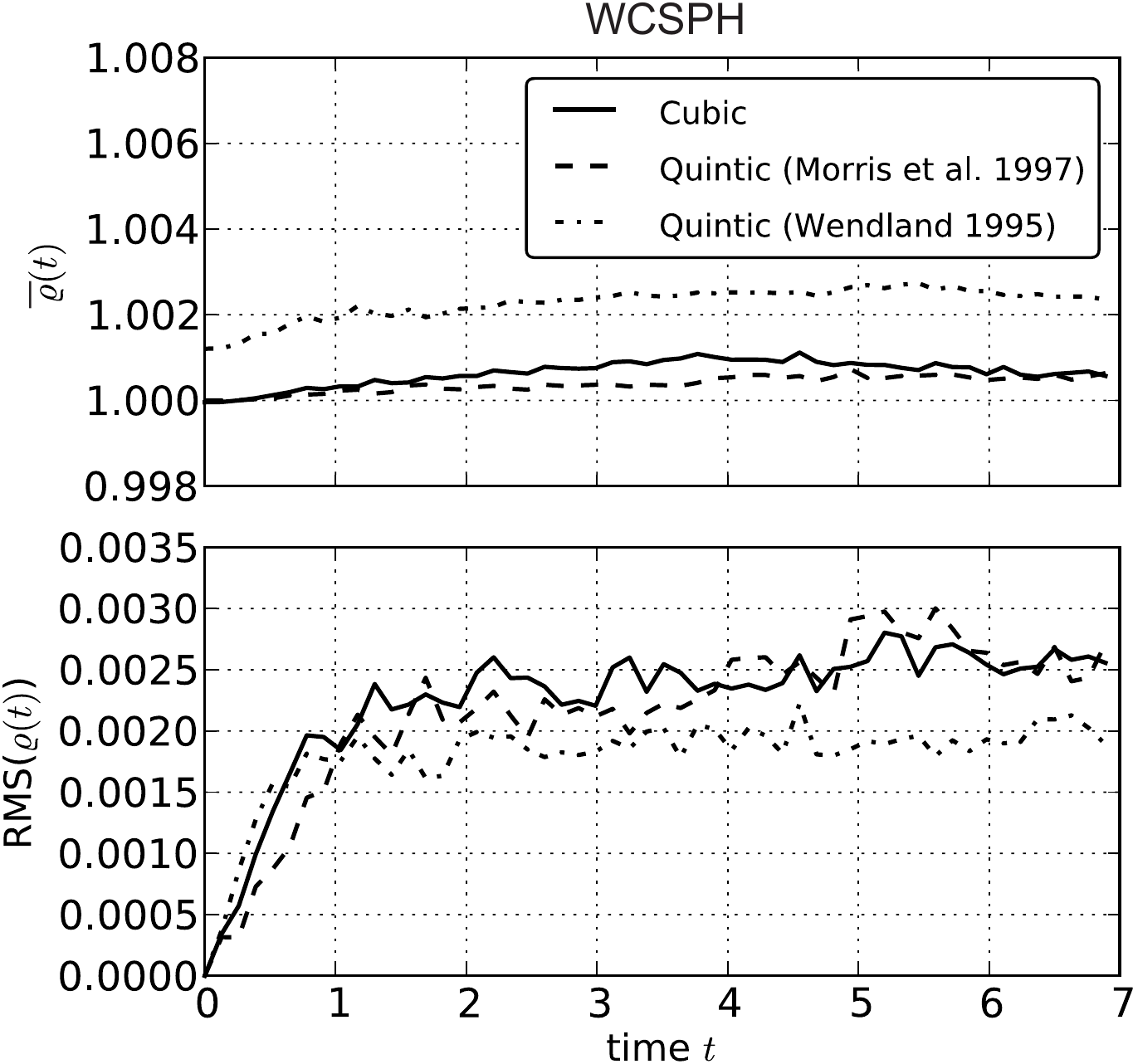} &
(b)
\includegraphics[width=0.45\textwidth]{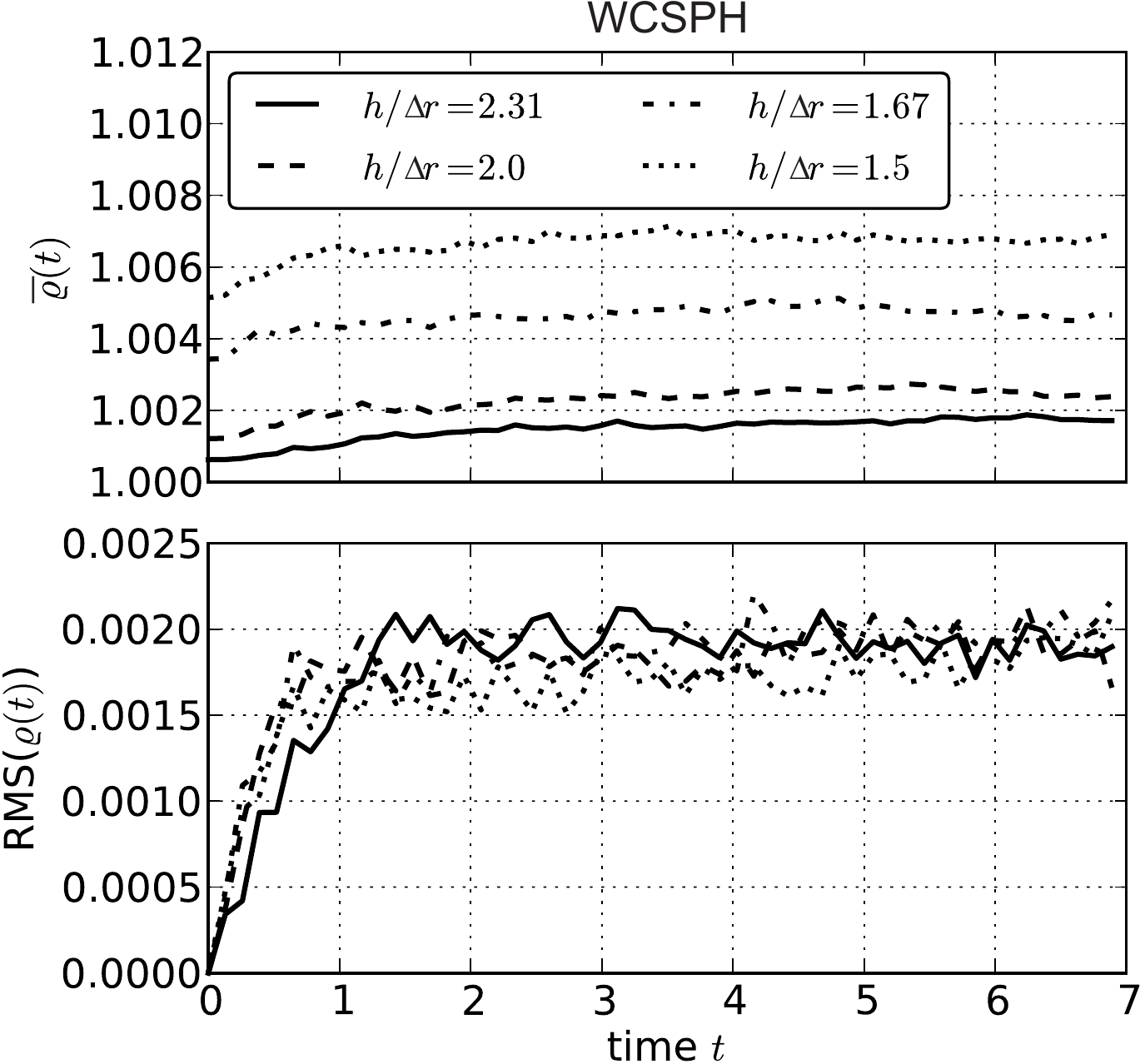} \vspace{20pt} \\
(c)
\includegraphics[width=0.45\textwidth]{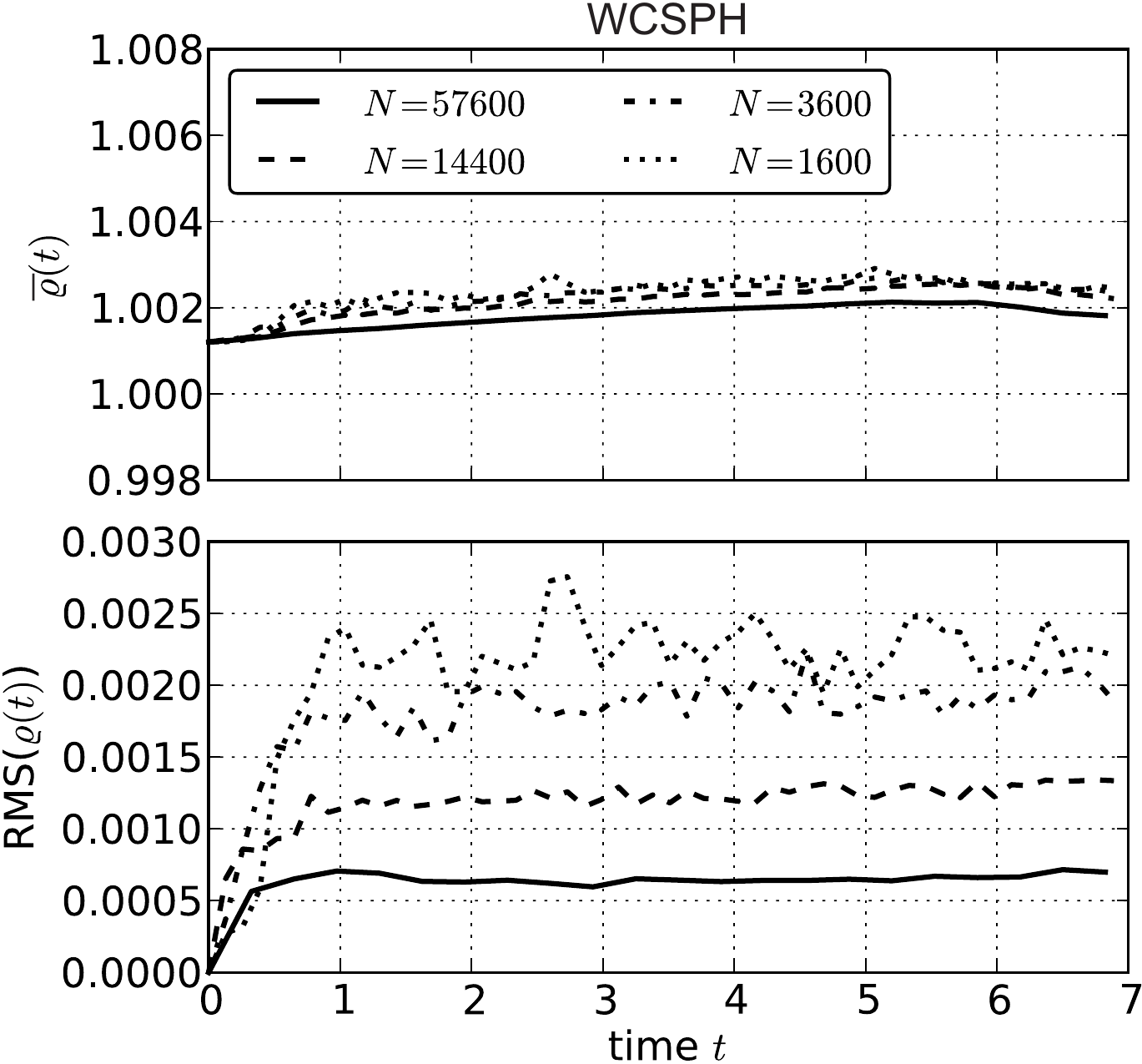} &
(d)
\includegraphics[width=0.45\textwidth]{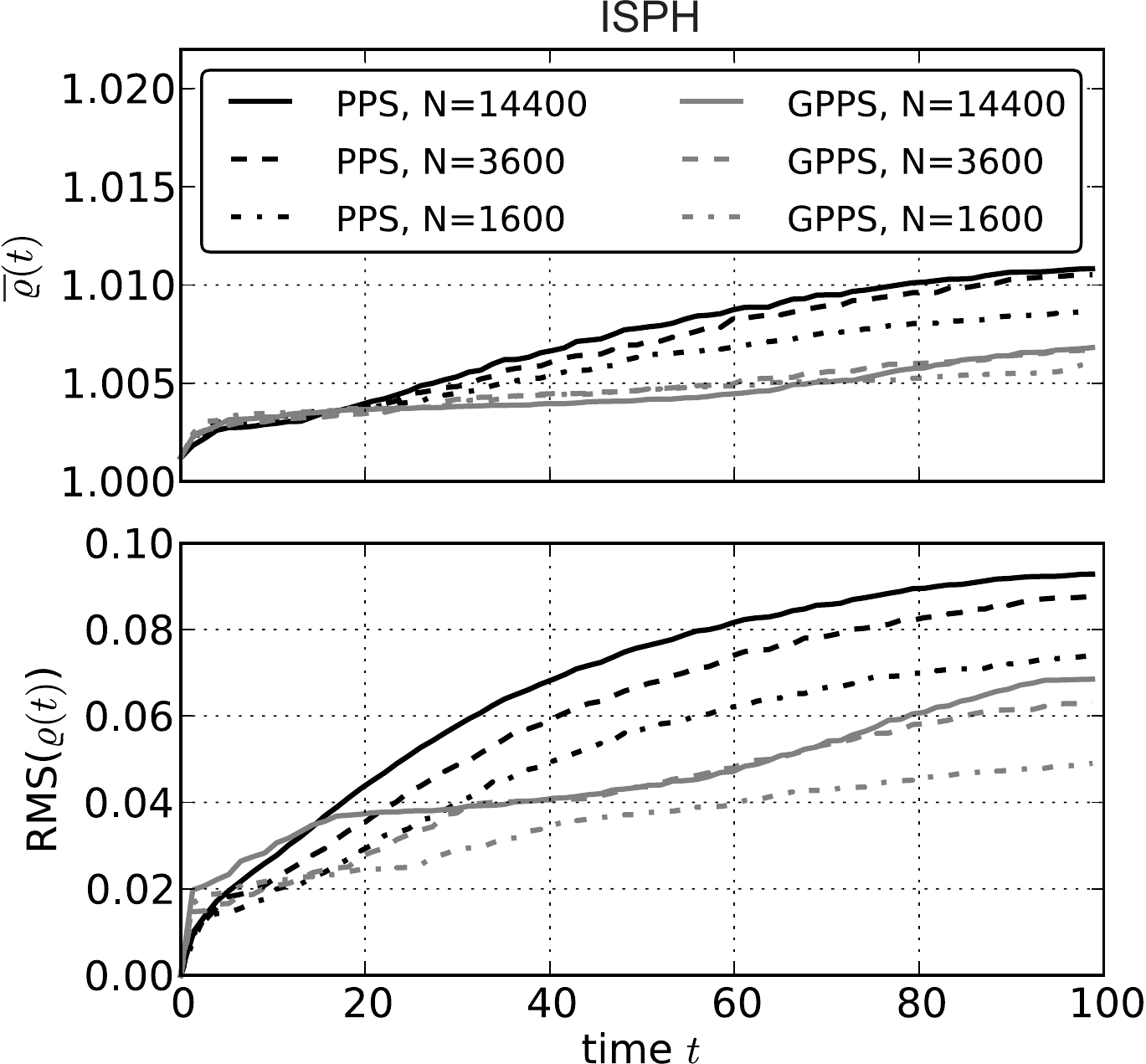}
\end{tabular}
\caption{The density mean value and the r.m.s.\ obtained for the lid-driven cavity flow at $\Reynolds=1000$; the effect of: (a) the kernel choice, (b) $h/\Delta r$, (c) number of particles $N$ influence in the WCSPH approach; (d) particles number $N$ influence in both: ISPH-PPS and ISPH-GPPS techniques.}
\label{fig:wcsph-lid-rms}
\end{figure}

In WCSPH, comparing the influence of the kernel shape, we observe that the quintic kernel proposed by Wendland (\ref{quintic Wendland}) gives the smallest fluctuations of the density field, cf. Fig.~\ref{fig:wcsph-lid-rms}(a). 
The weakness of the Wendland kernel is a slight overestimation of the mean density. Even for a homogeneously distributed set of particles ($t=0$) the density field is flawed, cf. discussion in \cite{Hongbin & Xin 2005}. At decrease of the $h/\Delta r$ parameter causes this error to grow from $0.2\%$ for $h/\Delta r=2.31$ up to about $0.8\%$ for $h/\Delta r=1.5$, cf. Fig.~\ref{fig:wcsph-lid-rms}(b). Interestingly, parameter $h/\Delta r$ has no significant effect on $\varrho_{\text{rms}}(t)$.
The particle number influence obtained with WCSPH is presented in Fig.~\ref{fig:wcsph-lid-rms}(c).
The density r.m.s.\, independently on the number of particles in domain, stabilizes after about $t=1$ and remains less than $1\%$ of initial density. Moreover, with increasing $N$ the density r.m.s.\ goes down.  
In the case of ISPH solvers, the growth of the r.m.s.\ stops only after obtaining 
the steady-state solution (roughly at $t=120$), cf. Fig.~\ref{fig:wcsph-lid-rms}(d). 
Intriguing is the fact that for the ISPH approach, the increase of the number of particles 
$N$ increases the r.m.s.  For ISPH-PPS the r.m.s.\ of the steady-state solution changes: 
from $7\%$ for $N=1600$ up to $9\%$ for $N=14400$. 
For ISPH-GPPS the density field is additionally affected by the projections 
between the particles and the grid. 
These projections cause additional smoothing and the density field r.m.s.\ of the steady-state solution varies: for $N=14400$ about $7\%$, for $N=3600$ about $6\%$ and $N=1600$ about $5\%$ of initial density.

\begin{figure}
\centering
\begin{tabular}{cc}
\includegraphics[width=0.48\textwidth]{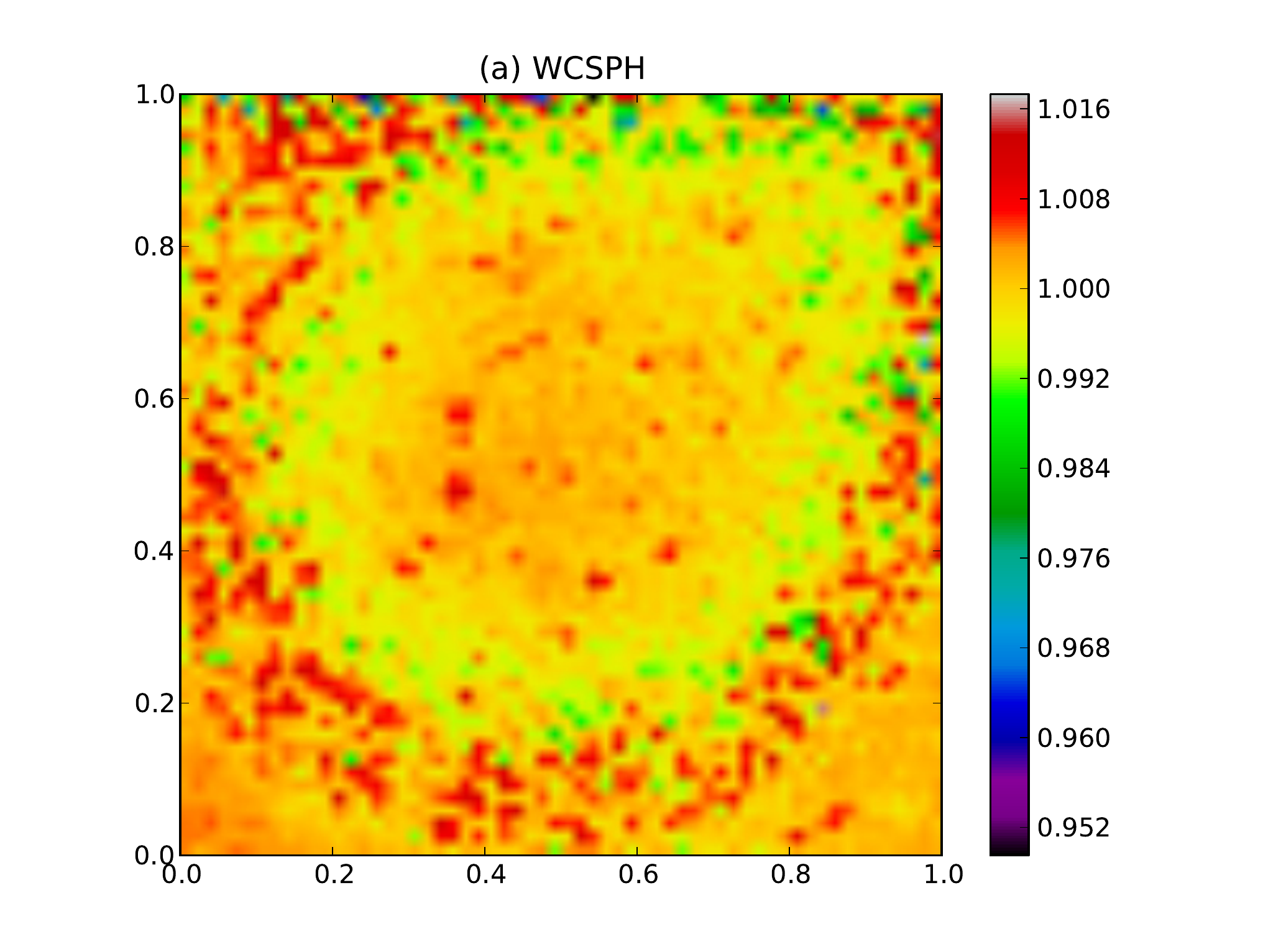}
\includegraphics[width=0.48\textwidth]{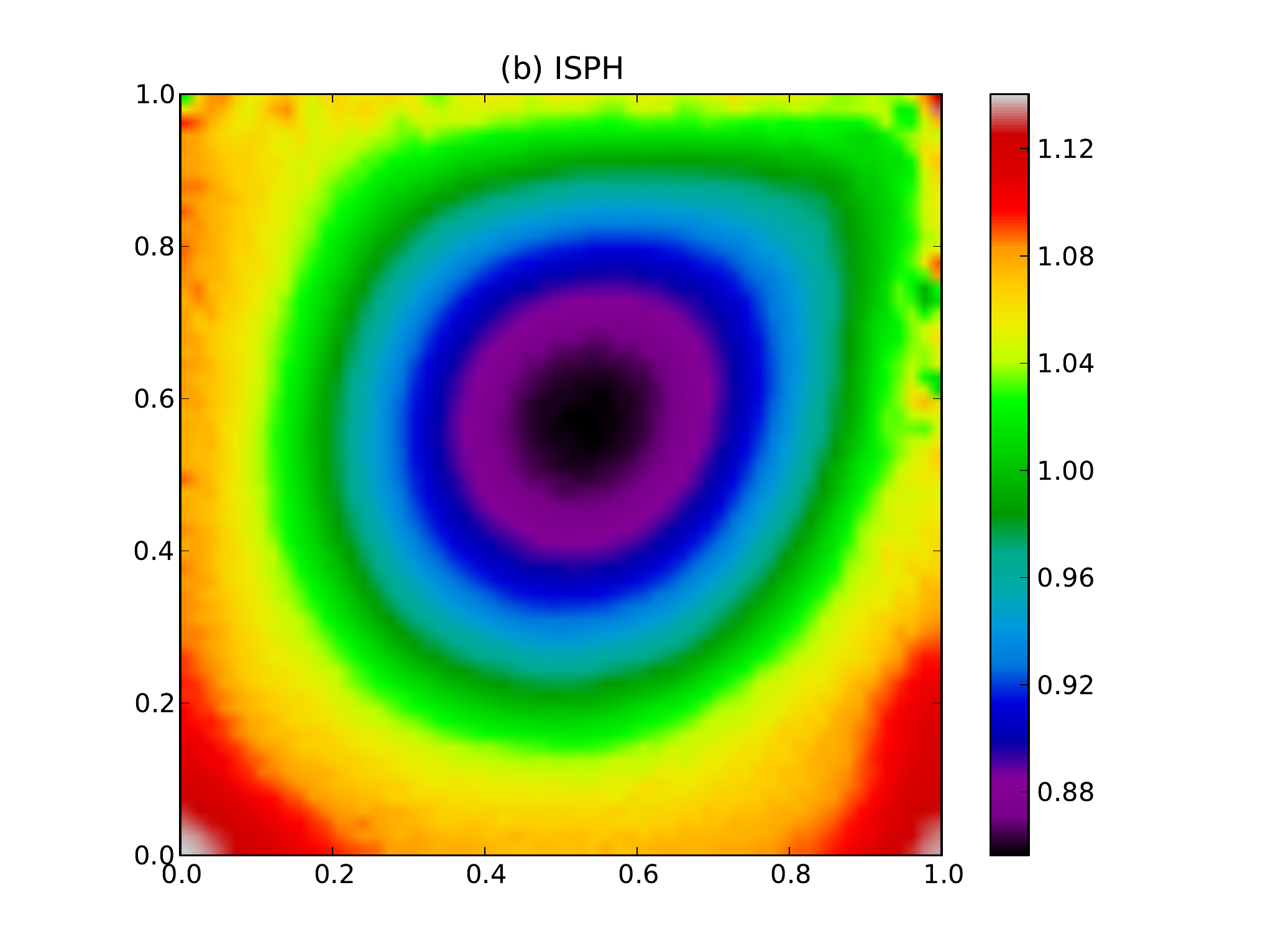}
\end{tabular}
\caption{The lid-driven cavity flow: density field at the steady-state solution.}
\label{fig:density-f}
\end{figure}
\begin{figure}
\centering
\begin{tabular}{ccc}
 \includegraphics[width=0.45\textwidth]{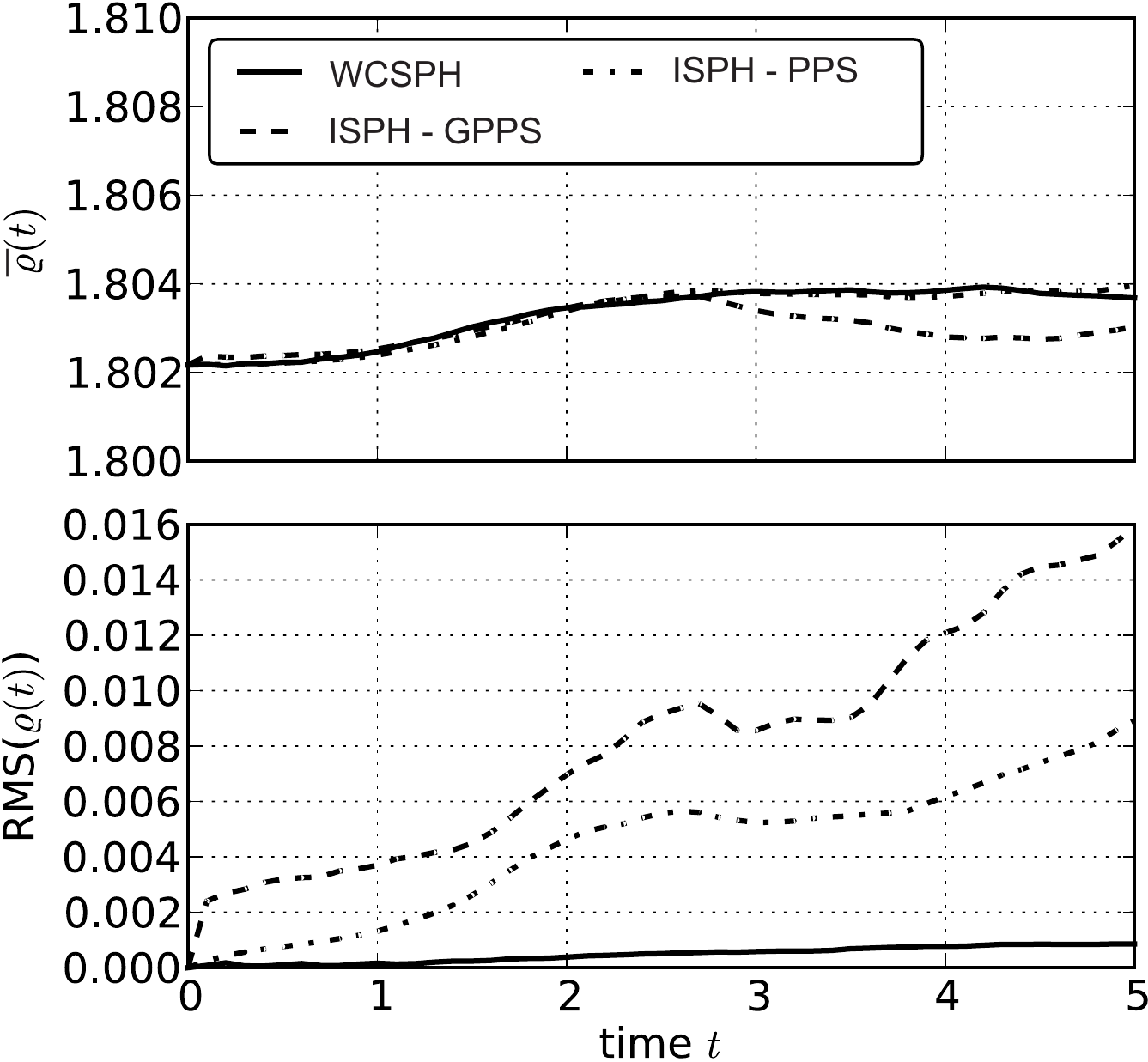}
\end{tabular}
\caption{The density mean value and the r.m.s.\ obtained for the Rayleigh-Taylor instability (only upper phase $\varrho_U=1.8\varrho_0$) using WCSPH and both ISPH approaches.}
\label{fig:density-rt}
\end{figure}

This discrepancy between WCSPH and ISPH density error values is caused by the completely different mechanisms of computing the dynamical pressure in the Navier-Stokes equation. Since in the case of the WCSPH approach, the suitably-chosen equation of state is used, the density field suffers from short-scale density waves presented in Fig.~\ref{fig:density-f}(a). In the case of ISPH, there is no explicit correction to the density. Therefore, during the computations, the density field strongly deviates from the level of the initial state, cf. Fig.~\ref{fig:density-f}(b). Since the Projection Methods give such considerable density errors, it is meaningful to implement and an additional correction procedure, cf. Sect.~\ref{sec:pozorski and wawrenczuk constant-density approach}.

In the case of the Rayleigh-Taylor instability, the comparison of density errors (cf. Fig.~\ref{fig:density-rt}) shows similar behavior as in the lid-driven cavity case.

\section{Constant-density ISPH approach}
\label{sec:pozorski and wawrenczuk constant-density approach}
In the previous section
we have shown that in the ISPH approach the density errors
cumulate during computations. To solve this problem, Pozorski and Wawre\'nczuk
\cite{Pozorski & Wawrenczuk 2002} suggested to use the second correction term
(a similar approach was later proposed, in the multi-phase flow context, by Hu and Adams \cite{Hu & Adams 2007}).
Let us consider incompressible fluid with initially uniform density $\varrho_0$.
After the time step performed with the ISPH approach, the corrected velocity
field satisfies the divergence-free condition; however, this procedure does not
explicitly guarantee that $\varrho = const$. The Pozorski and Wawre\'nczuk idea
consists in performing an additional correction to the particle positions
\begin{equation}
\mathbf r_a^{n+1} = \tilde{\mathbf r}_a^{n+1} - \frac{1}{\varrho_0} \nabla p_a^* = \mathbf r_a^n + \mathbf u_a^{n+1} \delta t - \frac{1}{\varrho_0} \nabla p_a^*,
\end{equation}
where $\mathbf u_a^{n+1}$ and $\tilde{\mathbf r}_a^{n+1}$ are respectively the
divergence-free velocity field and particle positions obtained after the ISPH
time step, Eqs.~(\ref{ISPH corrector}) and (\ref{Poisson}), while $p_a^*$ appears as a potential correction field computed from
\begin{equation} \label{Pozorski PPS}
\frac{1}{\varrho_0} \nabla \cdot \left( \frac{\varrho^n}{\varrho_0} \nabla p^* \right) = 1 - \frac{\tilde{\varrho}^{n+1}}{\varrho_0},
\end{equation}
where $\tilde{\varrho}^{n+1}$ is the density obtained after the ISPH time step. Both $\varrho^n$ and
$\tilde{\varrho}^{n+1}$ are computed using Eq.~(\ref{SPH direct density
computation multiphase}).
The Poisson Eq.~(\ref{Pozorski PPS}) is obtained from the request
\begin{equation}
  \varrho^{n+1}(\mathbf r_a) = m_a \sum_b W(\mathbf r_a - \mathbf r^{n+1}_b, h) = m_a \sum_b W(\mathbf r_a - \tilde{\mathbf r}_b^{n+1} - \nabla p_b^*, h) = \varrho_0(\mathbf r_a)
\end{equation}
by the Taylor series expansion around $\mathbf r_a - \tilde{\mathbf r}_b^{n+1}$.

\begin{figure}
\centering
\begin{tabular}{cc}
 \includegraphics[width=0.49\textwidth]{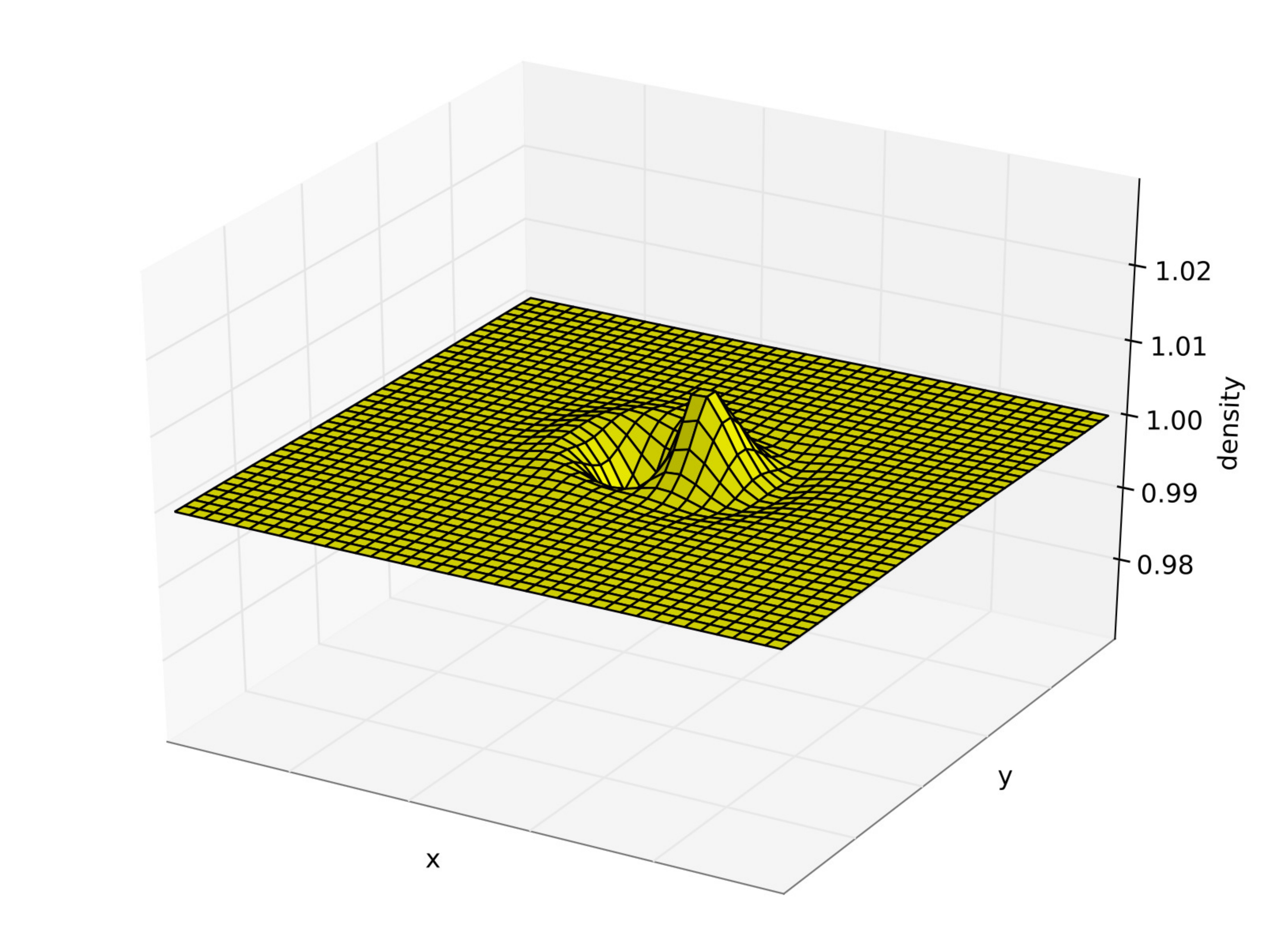} &
 \includegraphics[width=0.49\textwidth]{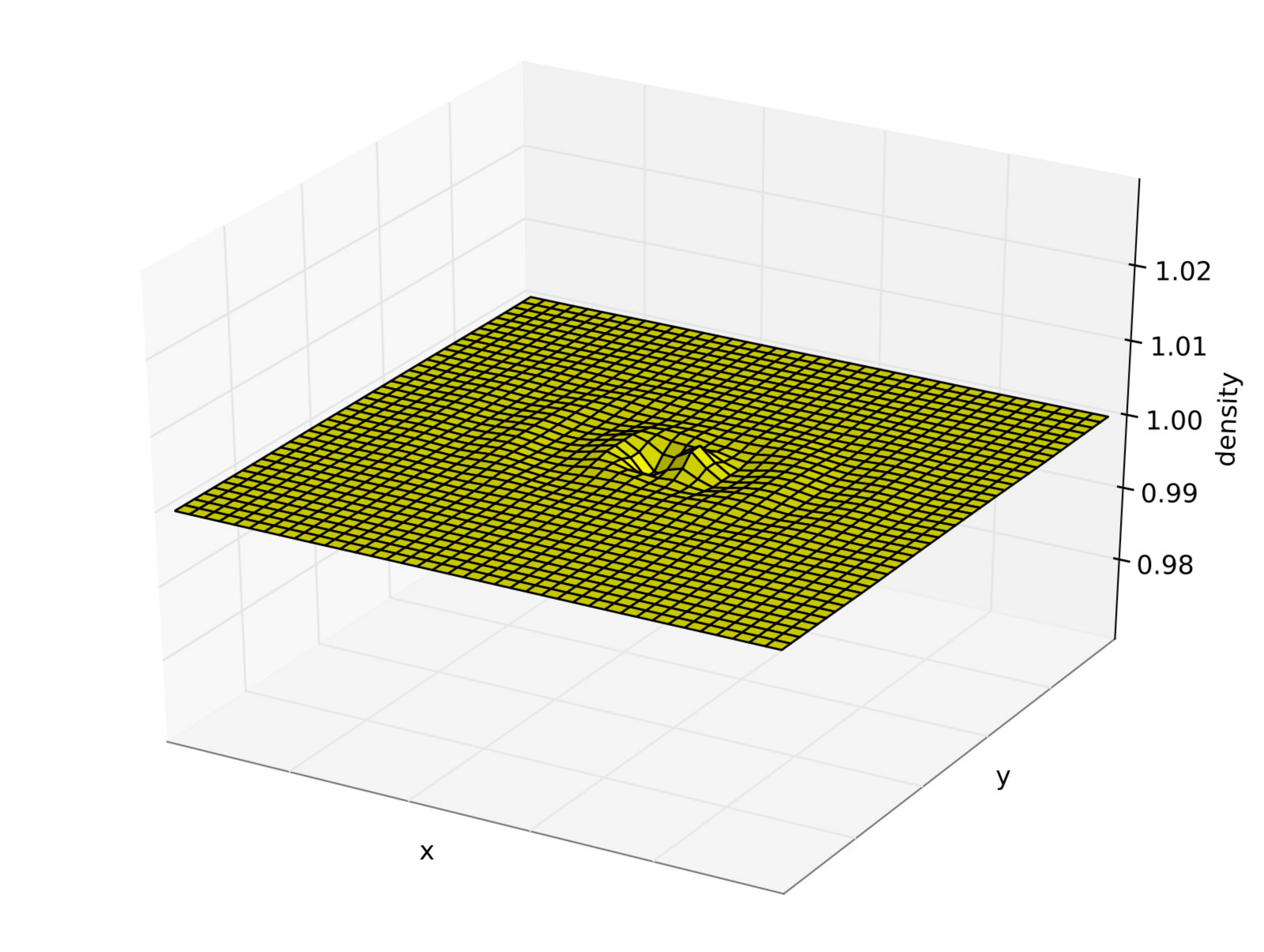}
\end{tabular}
\caption{The density field with local disturbance; left: the initial state
(regular set of particles with one particle displaced), right: after 50
correction iterations.}
\label{fig:poz-corr}
\end{figure}

Since, performing such a correction, the second Poisson equation has to be
solved, the computational effort is increased. However, there is no necessity to
compute the correction at each time step; rather, it is
applied only if the density error exceeds a certain threshold value. On the other hand,
the procedure may be performed several times in one time step.
Figure~\ref{fig:poz-corr} shows how the initial disturbance of the density field
(a regular set of particles with one particle displaced) is corrected 
with~(\ref{Pozorski PPS}).

\begin{figure}
\centering
\begin{tabular}{ccc}
\includegraphics[width=0.95\textwidth]{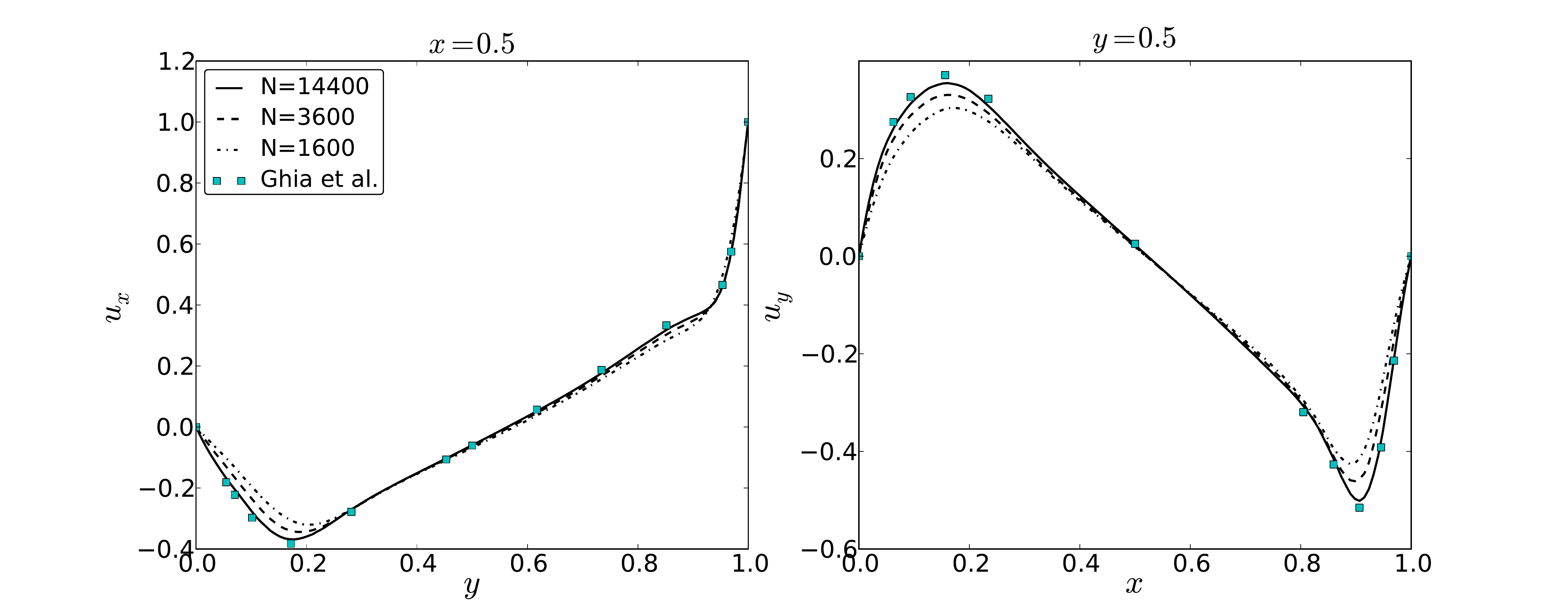} 
\end{tabular}
\caption{The lid-driven cavity velocity profiles at the steady-state against Ghia et al.~\cite{Ghia et al. 1982} reference data ($\Reynolds=1000$); results obtained using ISPH with PPS and Pozorski \& Wawre\'nczuk correction \cite{Pozorski & Wawrenczuk 2002} for the Wendland kernel~\cite{Wendland 1995}, $h/\Delta r = 2$ and different number of particles $N$.}
\label{fig:lid-poz}
\end{figure}

\begin{figure}
\centering
\begin{tabular}{cc}
\includegraphics[width=0.45\textwidth]{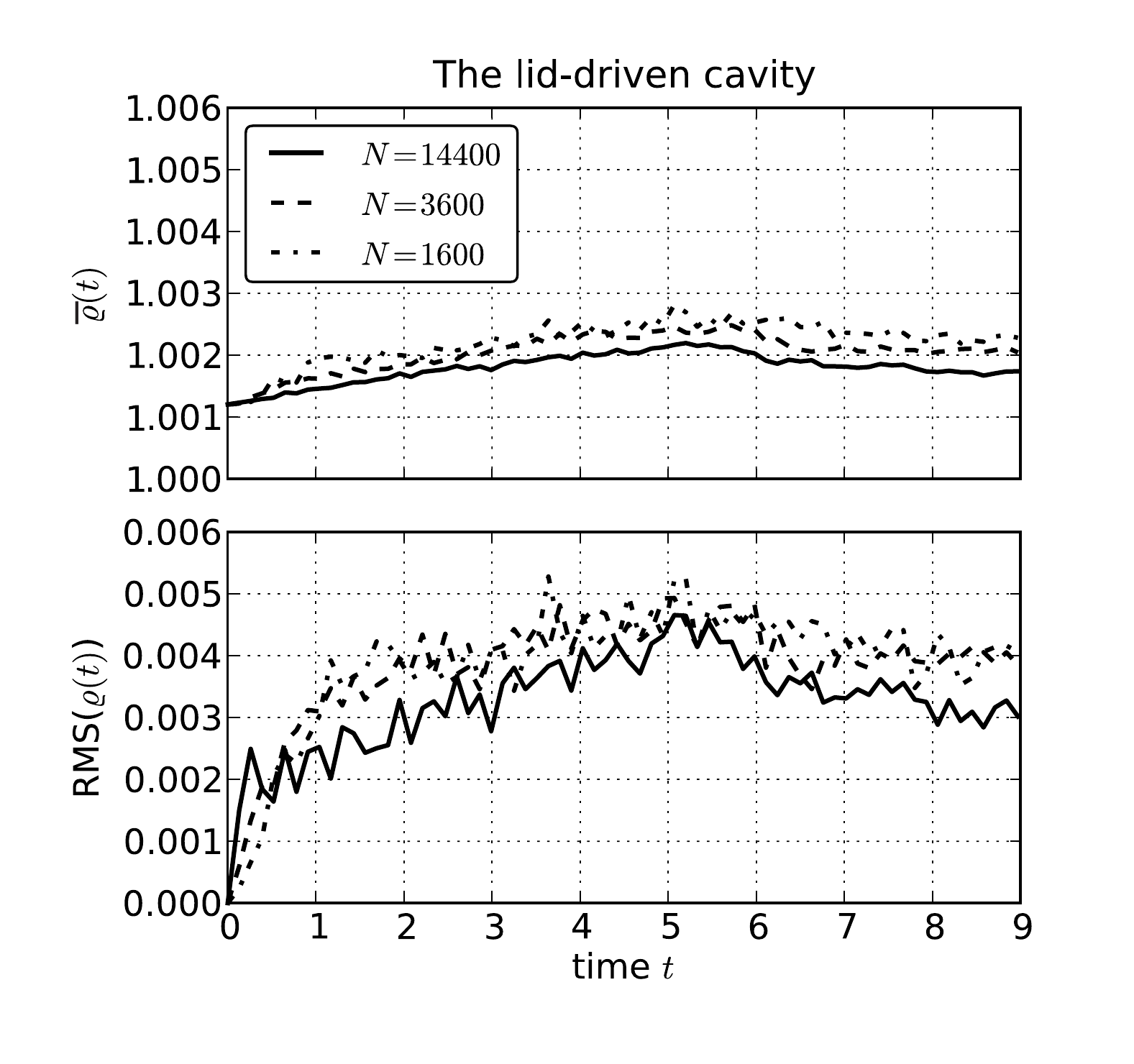}
\includegraphics[width=0.45\textwidth]{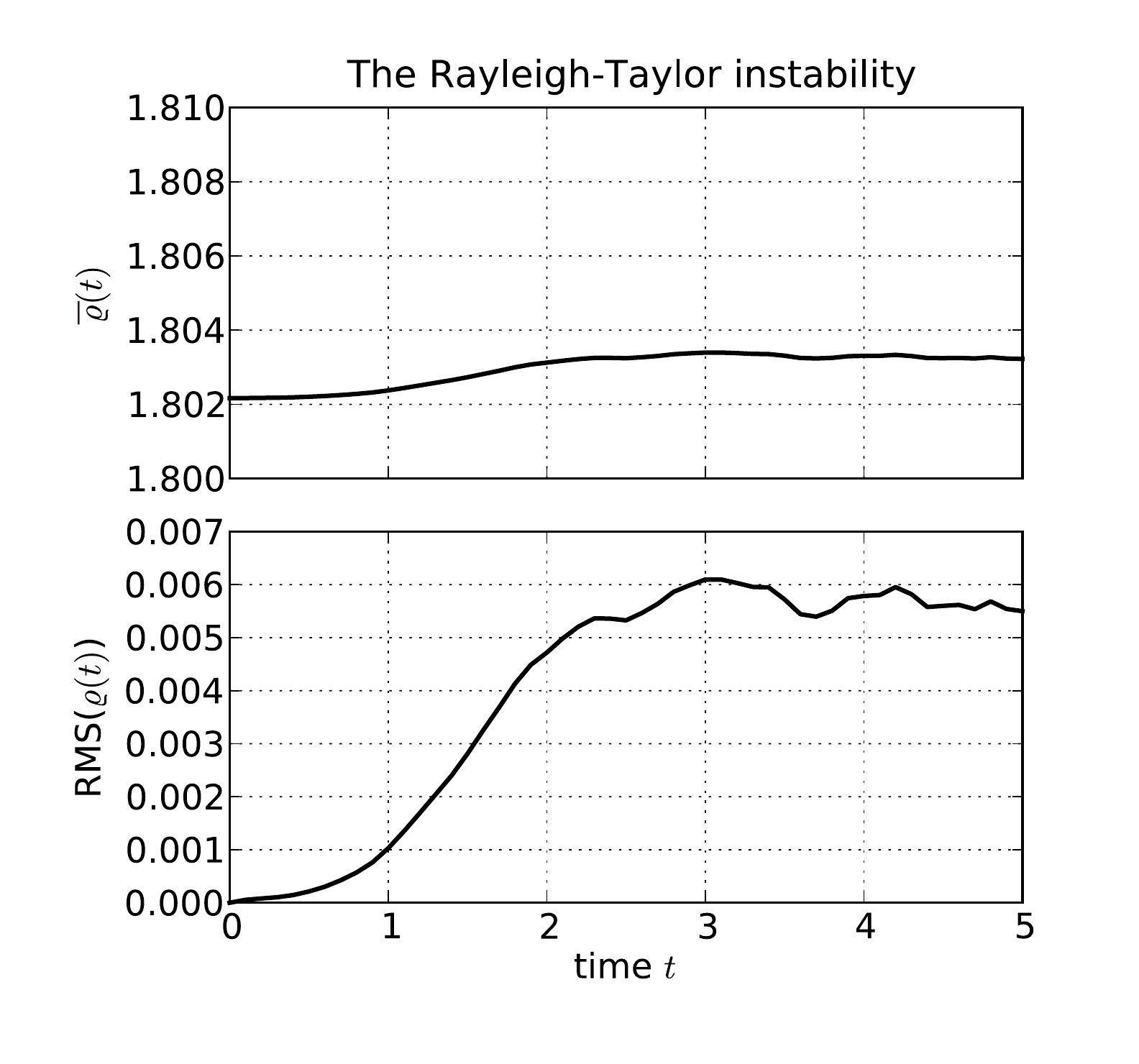}
\end{tabular}
\caption{The density mean value and r.m.s.\ obtained using the ISPH approach with PPS and Pozorski \& Wawre\'nczuk density correction \cite{Pozorski & Wawrenczuk 2002}; (a) the lid driven cavity ($\Reynolds=1000$), (b) the Rayleigh-Taylor instability ($\Reynolds=420$); the results obtained for the Wendland kernel~\cite{Wendland 1995} and $h/\Delta r = 2$.}
\label{fig:density-poz}
\end{figure}

\begin{figure}
\centering
\begin{tabular}{ll}
(a) & (b) \\
\includegraphics[width=0.49\textwidth]{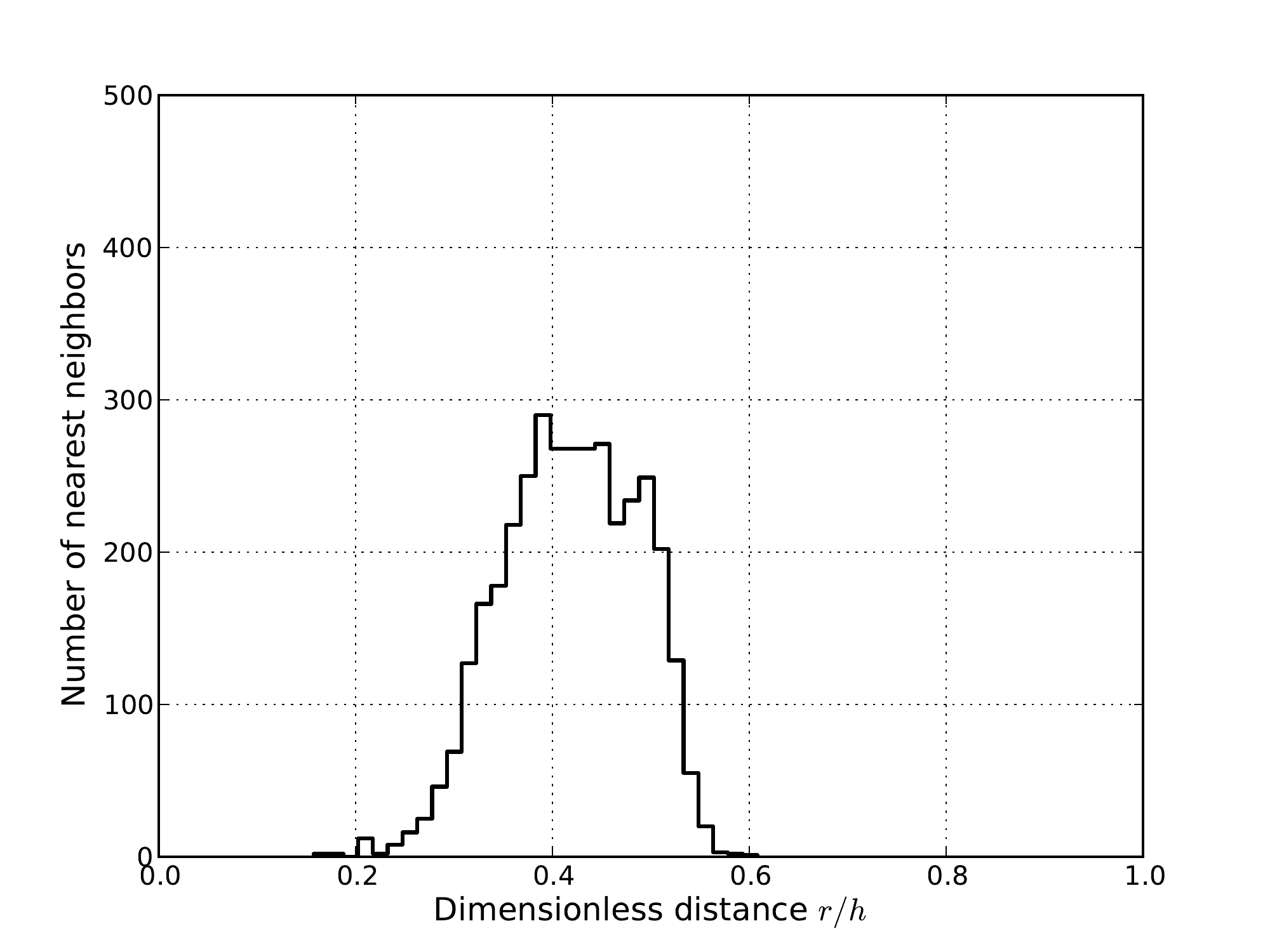} &
\includegraphics[width=0.49\textwidth]{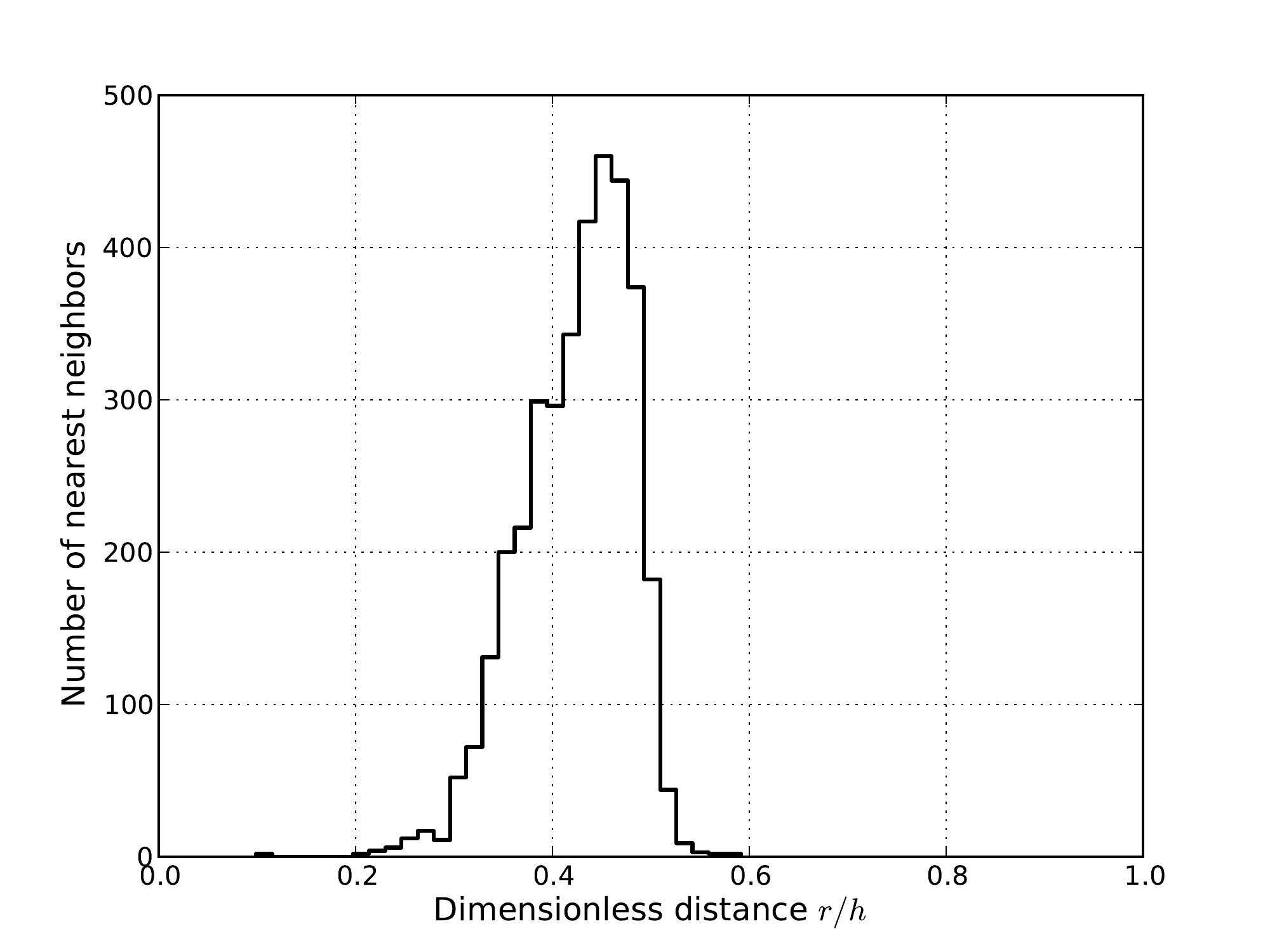}
\end{tabular}
\caption{Histograms of the distance between the nearest pairs of the particles; 
the results obtained for the ISPH-PPS approach (a) without and (b) with 
the Pozorski and Wawre\'nczuk correction procedure.}
\label{fig:histcor}
\end{figure}

To perform the test problems, we decided to utilize ISPH with particle
positions correction performed once per time step. 
Since both Poisson equations were solved on particles, we got fully grid-free approach. 
The velocity profiles of the lid-driven cavity problem ($\Reynolds=1000$) are presented in
Fig.~\ref{fig:lid-poz}. There are no noticeable differences between the velocities
obtained using ISPH solver with (Fig.~\ref{fig:lid-poz}) and without (Fig.~\ref{fig:lid-sph-n}) the
Pozorski and Wawre\'nczuk correction. However, comparing the density fields,
cf. Fig.~\ref{fig:density-poz}, with the results obtained without the correction,
cf. Fig.~\ref{fig:density-f}(d), the advantages of this approach become obvious: the
growth of the r.m.s.\ density stops after $t=5$ and stays at a level less
than 0.1\%. 
The regularisation of the particles' distribution is also visible in the histograms of 
the distance to the nearest neighbor of each particle: the comparison of histograms for ISPH-PPS with
and without the correction procedure is presented in Fig.~\ref{fig:histcor}.
Another convincing argument for the utility of the Pozorski and
Wawre\'nczuk approach is the computational effort. Since, as in the case of
WCSPH, the density error does not accumulate, and, despite the use of
two Poisson solvers, the computational time is about 3 times shorter (and about 2 times longer than ISPH-PPS), 
the use of the density correction algorithm seems to be profitable.

On the other hand, it is interesting to note a higher convergence rate with increasing number of particles in domain, 
as compared to the solutions obtained by Lee et al.~\cite{Lee et al. 2008} (Fig. 6) and Xu et al.~\cite{Xu et al. 2009} (Fig. 25).
We suppose that a better convergence rate in our case is due to a proper choice of computational parameters ($h/\Delta r$, kernel type and b.c.).

\begin{figure}
\centering
\begin{tabular}{ccc}
 \includegraphics[width=0.33\textwidth]{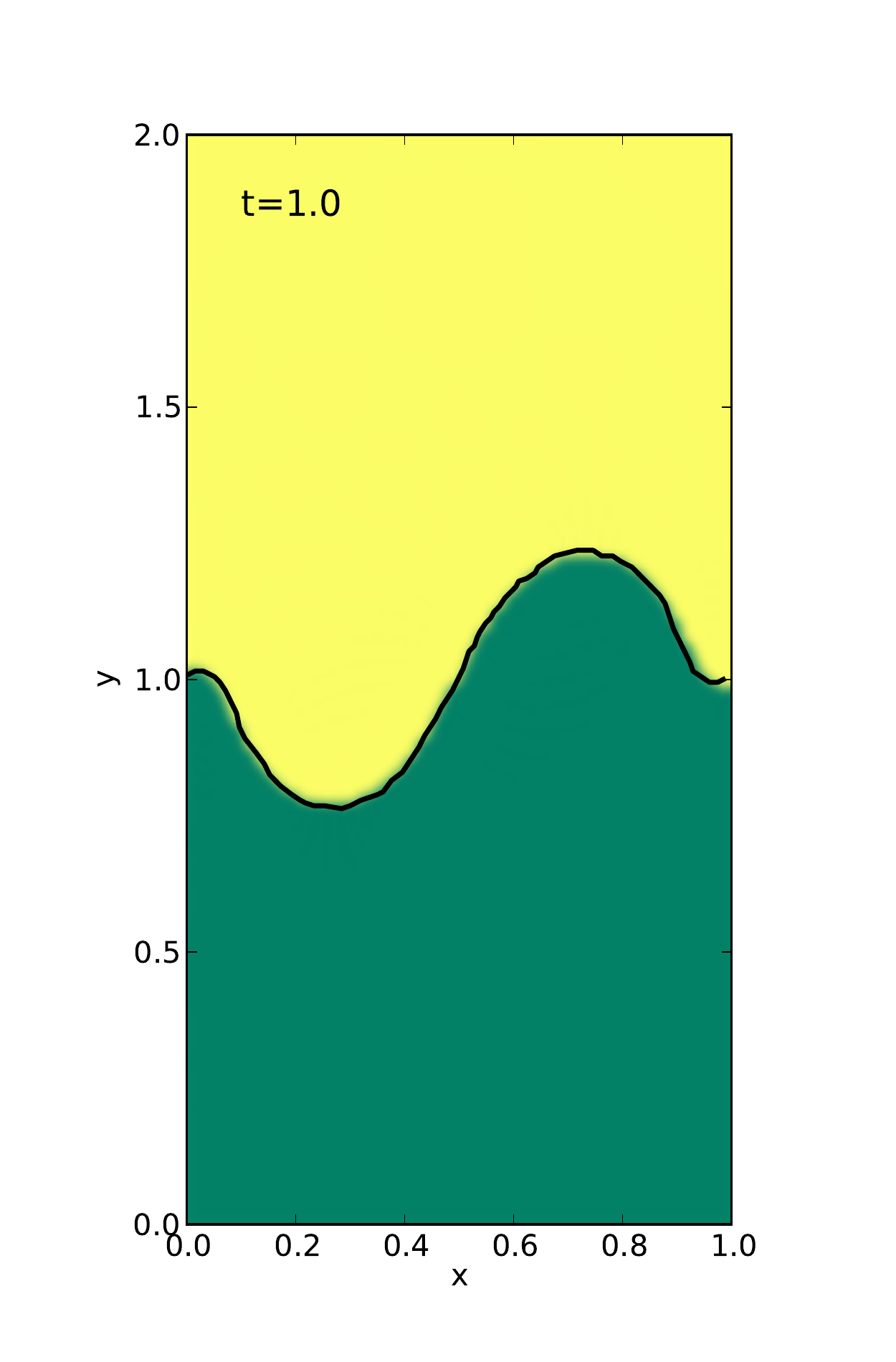}
 \includegraphics[width=0.33\textwidth]{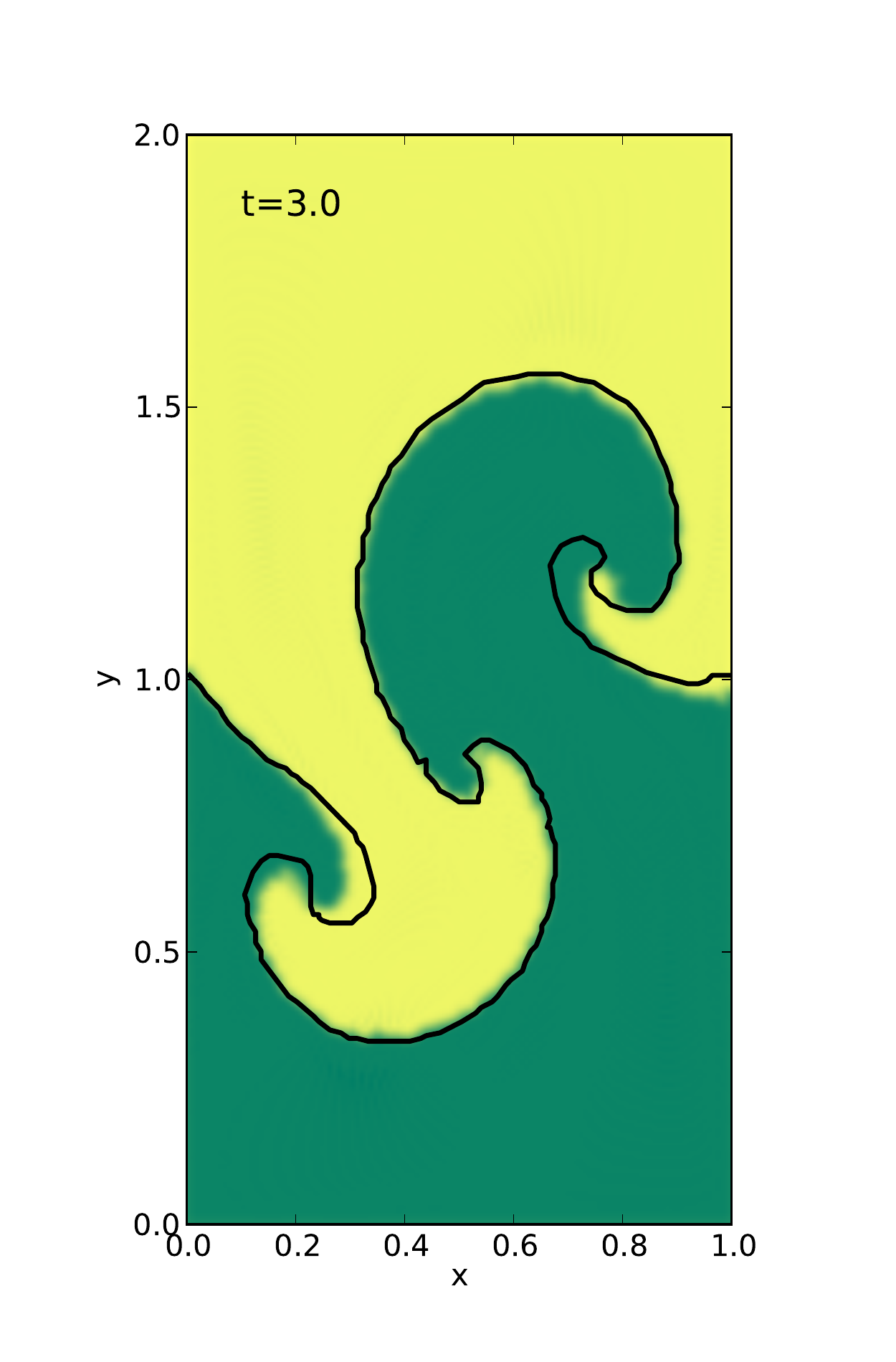}
 \includegraphics[width=0.33\textwidth]{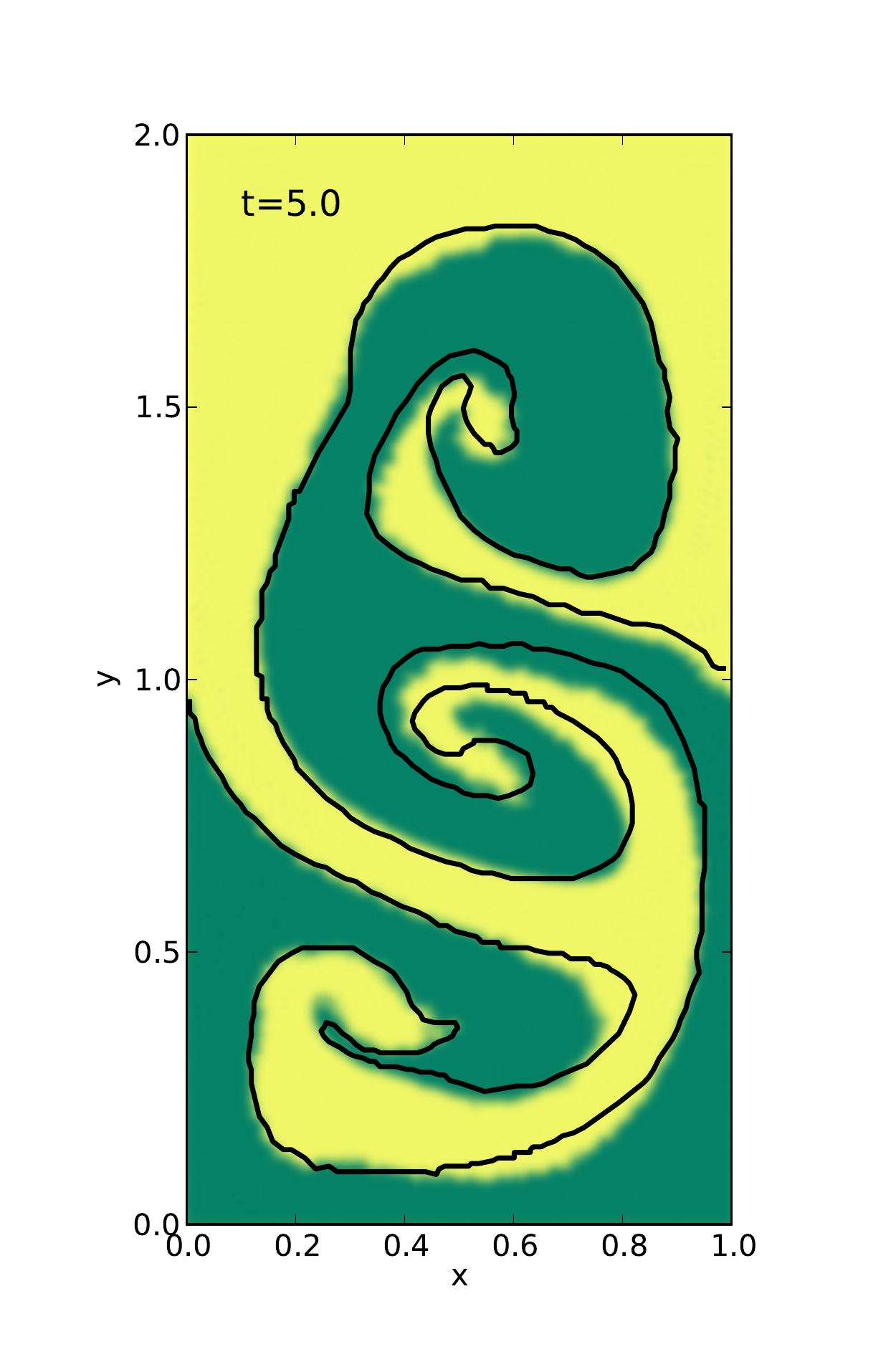}
\end{tabular}
\caption{The Rayleigh-Taylor instability ($\Reynolds=420$) computed using ISPH with PPS and Pozorski \& Wawre\'nczuk density correction \cite{Pozorski & Wawrenczuk 2002}; the black lines denote the interface position obtained by Grenier et al.~\cite{Grenier et al. 2009} using the Level-Set formulation ($312\times 624$ cells).}
\label{fig:rt-poz}
\end{figure}

For the Rayleigh-Taylor instability, the calculated particle positions at times
$t=1$, $3$ and $5$ are presented in Fig.~\ref{fig:rt-poz}. The comparison with
the reference interface shapes \cite{Grenier et al. 2009} show quite a good
agreement. What is more, there are no considerable differences between ISPH with
and without the Pozorski and Wawre\'nczuk correction term. Moreover, as in the
case of the lid-driven cavity, the use of such a correction term
assures that the density error (Fig.~\ref{fig:density-poz}) is kept below the desired level and
the computational time is still acceptable (in comparison with the WCSPH approach).

\section{Conclusions}
\label{sec:conclusions}

In this paper three different SPH incompressibility treatments were
considered. To validate these approaches, two tests cases were
simulated: the lid-driven cavity at $\Reynolds=1000$
and the Rayleigh-Taylor instability at $\Reynolds=420$. 
Summarising, in comparison to the previous tests presented by Lee et al.~\cite{Lee et al. 2008} and Xu et al.~\cite{Xu et al. 2009},
in all of the SPH incompressibility treatments, the velocity field exhibit a better convergence rate, when the number of particles in the domain grows.
Detailed study of the incompressibility treatments revealed some pros and cons of the approaches. The main
disadvantage of the WCSPH approach is its high computational cost. Therefore,
from this point of view, it is better to use the ISPH approaches. Nevertheless,
both the ISPH-GPPS and the ISPH-PPS solvers in their original formulation suffer from the density error
accumulation. 

For the fluid flows where besides the properly modelled
velocity field, the correct density field is of importance, an interesting
alternative to WCSPH is the ISPH approach with the separate density correction. 
In this approach the density accumulation error can be reduced
to a specified level while still retaining the grid-free formulation. 
It is important to note that, as such, none of the ISPH
approaches can be used to properly model gas-liquid flows such as a
bubble raising in water, where the divergence-free condition is
satisfied only for the liquid phase. Such flows can be relatively easily modelled
with the WCSPH approach, where each phase can be computed with a different
equation of state. 

The study has also shown the importance of correct choice of computational parameters:
the kernel formula $W$, its smoothing length $h$ for given inter-particle distance 
$h/\Delta r$, and the number of particles $N$ in the domain. The impact of those 
parameters on the results has been analyzed and optimal kernel has been found (the Wendland formula).
Regarding the other quantities, they represent a compromise between the CPU efficiency and accuracy.
 Further work will be undertaken to extend
the ISPH approach making it possible to model gas-liquid mixed flows with the
interface phenomena such as the surface tension or the Marangoni effect. 
Also, there is the need of appropriate wall b.c.\ for SPH simulations in complex geometries (arbitrary shape of the boundary).

 

\newpage
\section*{List of figure captions}

Fig.\ 1: The ghost-particle no-slip boundary scheme: a) straight wall, b) inner corner.

Fig.\ 2: The problem with the ghost-particle no-slip boundary condition near the corner: depending on value of the parameter $\alpha$, the velocity boundary condition changes.

Fig.\ 3: The WCSPH results of the lid-driven cavity ($\Reynolds=1000$) at $t=0.03$ with: a) the no-slip and b) the free-slip boundary treatment for velocity divergence computation; employing no-slip condition induces instabilities near the corners.

Fig.\ 4: The ghost-particle free-slip boundary treatment for: a) the straight wall and b) the corner.

Fig.\ 5: The lid-driven cavity steady-state velocity profiles for: (a) WCSPH, (b) ISPH-GPPS and (c) ISPH-PPS against Ghia et al.~\cite{Ghia et al. 1982} results; profiles obtained for different kernels; $N=3600$, $h/\Delta r = 2$.

Fig.\ 6: The lid-driven cavity steady-state solution ($\Reynolds=1000$, $N=3600$) computed with the WCSPH approach and kernels: (\ref{cubic spline kernel}), (\ref{quintic Wendland}), (\ref{quintic Morris}); the particle clustering phenomenon is noticeable with the cubic spline kernel.

Fig.\ 7: Histograms of the distance between the nearest pairs of particles; 
the results obtained for the WCSPH approach and kernels: (\ref{cubic spline kernel}), 
(\ref{quintic Wendland}), (\ref{quintic Morris}).

Fig.\ 8: The lid-driven cavity steady-state velocity profiles for: (a) WCSPH, (b) ISPH-GPPS and (c) ISPH-PPS against Ghia et al.~\cite{Ghia et al. 1982} results; profiles obtained with different $h/\Delta r$ values; results obtained using the Wendland kernel~\cite{Wendland 1995} and $N=3600$ particles in domain.

Fig.\ 9: The lid-driven cavity steady-state velocity profiles for: (a) WCSPH, (b) ISPH-GPPS and (c) ISPH-PPS  against Ghia et al.~\cite{Ghia et al. 1982} results; results for different number of particles $N$; data obtained using the Wendland kernel~\cite{Wendland 1995} and $h/\Delta r = 2$.

Fig.\ 10: The CPU times to obtain the steady-state solution of the lid-driven cavity ($\Reynolds=1000$) using the WCSPH and both ISPH approaches; for the WCSPH method, two continuity equations are compared: Eq.~(\ref{SPH continuity symmetrical}) and density definition (\ref{SPH direct density computation multiphase}).

Fig.\ 11: The Rayleigh-Taylor instability; particle positions at $t=5$ obtained using different incompressibility treatments; solid line: liquid-liquid interface from the reference Level-Set solution \cite{Grenier et al. 2009}.

Fig.\ 12: The density mean value and the r.m.s.\ obtained for the lid-driven cavity flow at $\Reynolds=1000$; the effect of: (a) the kernel choice, (b) $h/\Delta r$, (c) number of particles $N$ influence in the WCSPH approach; (d) particles number $N$ influence in both: ISPH-PPS and ISPH-GPPS techniques.

Fig.\ 13: The lid-driven cavity flow: density field at the steady-state solution.

Fig.\ 14: The density mean value and the r.m.s.\ obtained for the Rayleigh-Taylor instability (only upper phase $\varrho_U=1.8\varrho_0$) using WCSPH and both ISPH approaches.

Fig.\ 15: The density field with local disturbance; left: the initial state
(regular set of particles with one particle displaced), right: after 50
correction iterations.

Fig.\ 16: The lid-driven cavity velocity profiles at the steady-state against Ghia et al.~\cite{Ghia et al. 1982} reference data ($\Reynolds=1000$); results obtained using ISPH with PPS and Pozorski \& Wawre\'nczuk correction \cite{Pozorski & Wawrenczuk 2002} for the Wendland kernel~\cite{Wendland 1995}, $h/\Delta r = 2$ and different number of particles $N$.

Fig.\ 17: The density mean value and r.m.s.\ obtained using the ISPH approach with PPS and Pozorski \& Wawre\'nczuk density correction \cite{Pozorski & Wawrenczuk 2002}; (a) the lid driven cavity ($\Reynolds=1000$), (b) the Rayleigh-Taylor instability ($\Reynolds=420$); the results obtained for the Wendland kernel~\cite{Wendland 1995} and $h/\Delta r = 2$.

Fig.\ 18: Histograms of the distance between the nearest pairs of the particles; 
the results obtained for the ISPH-PPS approach (a) without and (b) with 
the Pozorski and Wawre\'nczuk correction procedure.

Fig.\ 19: The Rayleigh-Taylor instability ($\Reynolds=420$) computed using ISPH with PPS and Pozorski \& Wawre\'nczuk density correction \cite{Pozorski & Wawrenczuk 2002}; the black lines denote the interface position obtained by Grenier et al.~\cite{Grenier et al. 2009} using the Level-Set formulation ($312\times 624$ cells).

\begin{thebibliography}{}

\bibitem{Gingold & Monaghan 1977}
R.A. Gingold, J.J. Monaghan, Smoothed Particle Hydrodynamics: Theory and application to non-spherical stars, Mon. Not. R. Astron. Soc. 181 (1977) 375-389.

\bibitem{Lucy 1977}
L.B. Lucy, A numerical approach to the testing of the fission hypothesis, Astron. J. 82 (1977) 1013-1024.

\bibitem{Cummins & Rudman 1999}
S.J. Cummins, M. Rudman, An SPH projection method, J. Comput. Phys. 152 (1999) 584-607.


\bibitem{Lee et al. 2008}
E.-S. Lee, C. Moulinec, R. Xu, D. Violeau, D. Laurence, P. Stansby, Comparisons of weakly compressible and truly incompressible algorithms for the SPH mesh free particle method, J. Comput. Phys. 227 (2008) 8417-8436.

\bibitem{Xu et al. 2009}
R. Xu, P. Stansby, D. Laurence, Accuracy and stability in incompressible SPH (ISPH) based on the projection method and a new approach,
J. Comput. Phys. 228 (2009) 6703-6725.

\bibitem{Hu & Adams 2007}
X.Y. Hu, N.A. Adams, An incompressible multi-phase SPH method, J. Comput. Phys. 227 (2007) 264-278.

\bibitem{Pozorski & Wawrenczuk 2002}
J. Pozorski, A. Wawre\'nczuk, SPH computation of incompressible viscous flows, J. Theor. Appl. Mech. 40 (2002) 917-937.

\bibitem{Minier & Pozorski 1999}
J.-P. Minier, J. Pozorski, Wall boundary conditions in PDF methods and application to a turbulent flow, Phys. Fluids 11 (1999) 2632-2644.

\bibitem{Valizadeh et al. 2008}
A. Valizadeh, M. Shafieefar, J.J. Monaghan, S.A. Salehi Neyshaboori, Modeling two-phase flows using SPH method, J. Applied Sci. 8 (2008) 3817-3826.

\bibitem{Wendland 1995}
H. Wendland, Piecewise polynomial, positive definite and compactly supported radial functions of minimal degree, Adv. Comput. Math. 4 (1995) 389-396.

\bibitem{Morris et al. 1997}
J.P. Morris, P.J. Fox, Y. Zhu, Modeling low Reynolds number incompressible flows using SPH, J. Comput. Phys. 136 (1997) 214-226.

\bibitem{Feldman & Bonet 2007}
J. Feldman, J. Bonet, Dynamic refinement and boundary contact forces in SPH with applications in fluid flow problems, Int. J. Numer. Meth. Engng. 72 (2007) 295-324.

\bibitem{Morris 1996}
J.P. Morris, Analysis of Smoothed Particle Hydrodynamics with Applications, (doctor's thesis, Department of Mathematics, Monash University), 1996.

\bibitem{Hu & Adams 2006}
X.Y. Hu, N.A. Adams, A multi-phase SPH method for macroscopic and mesoscopic flows, J. Comput. Phys. 213 (2006) 844-861.

\bibitem{Colagrossi & Landrini 2003}
A. Colagrossi, M. Landrini, Numerical simulation of interfacial flows by smoothed particle hydrodynamics, J. Comput. Phys. 191 (2003) 227-264.

\bibitem{Cummins et al. 1997}
S.J. Cummins, M.J. Rudman, J.J. Monaghan, Projection methods and SPH, Monash University Applied Mechanics Report and Preprints, 1997.

\bibitem{Cleary & Monaghan 1999}
P.W. Cleary, J.J. Monaghan, Conduction modelling using smoothed particle hydrodynamics, J. Comput. Phys. 148 (1999) 227-264.

\bibitem{Batchelor 1967}
G.K. Batchelor, An Introduction to Fluid Dynamics, Cambridge Univ. Press, 1967.

\bibitem{Monaghan 1994}
J.J. Monaghan, Simulating free surface flows with SPH, J. Comput. Phys. 110 (1994) 399-406.

\bibitem{Gryffits 1999}
D.J. Gryffits, Introduction to Electrodynamics - 3rd ed., Prentice Hall, New Jersey, 1999.


\bibitem{Monaghan 1992}
J.J. Monaghan, Smoothed Particle Hydrodynamics, Annu. Rev. Astron. Astrophys. 30 (1992) 542-574.

\bibitem{Monaghan 1989}
J.J. Monaghan, On the problem of penetration in particle methods, J. Comput. Phys. 82 (1989) 1-15.

\bibitem{Campbell 1989}
P.M. Campbell, Some new algorithms for boundary value problems in smoothed particle hydrodynamics, Technical Report NA-TR-88-296, Mission Research Corporation, Albuquerque, 1989.

\bibitem{Shao & Lo 2003}
S. Shao, E.Y. Lo, Incompressible SPH method for simulating Newtonian and non-Newtonian flows with a free surface, Adv. Water Resour. 26 (2003) 787-800.

\bibitem{Yildiz et al. 2008}
M. Yildiz, R.A. Rook, A. Suleman, SPH with multiple boundary tangent method, Int. J. Numer. Meth. Engng. 77 (2009) 1416–1438.

\bibitem{Maxwell 1873}
J.C. Maxwell, A treatise on electricity and magnetism, Clarendon Press, Oxford, 1873.

\bibitem{Ghia et al. 1982}
U. Ghia, K.N. Ghia, C.T. Shin, High Re-solution for incompressible flow using the Navier-Stokes equations and a multigrid method, J. Comput. Phys. 48 (1982) 387-411.

\bibitem{Robinson 2009}
M. Robinson, Turbulence and Viscous Mixing using Smoothed Particle Hydrodynamics, PhD thesis, Monash University, Australia, 2009.

\bibitem{Szewc et al. 2011}
K. Szewc, J. Pozorski, A. Tani\`ere, Modeling of natural convection with Smoothed Particle Hydrodynamics: Non-Boussinesq formulation, Int. J. Heat Mass Transfer 54 (2011) 4807-4816.

\bibitem{Grenier et al. 2009}
N. Grenier, M. Antuono, A. Colagrossi, D. Le Touz\'e, B. Alessandrini, An Hamiltonian interface SPH formulation for multi-fluid and free-surface flows, J. Comput. Phys. 228 (2009) 8380-8393.

\bibitem{Hongbin & Xin 2005}
J. Hongbin, D. Xin, On criterions for smoothed particle hydrodynamics kernels in stable field, J. Comput. Phys. 202 (2005) 699-709.

\end{thebibliography}
\end{document}